\documentclass[twocolumn,english,aps,prb,showpacs,superscriptaddress,amssymb,amsfonts]{revtex4}
\usepackage{amsmath}
\usepackage{graphicx}
\usepackage{amssymb}
\usepackage{color}
\usepackage{tabularx}
\makeatletter
\newcommand{\be}{\begin{equation}}
\newcommand{\ee}{\end{equation}}
\newcommand{\bea}{\begin{eqnarray}}
\newcommand{\eea}{\end{eqnarray}}

\makeatother

\makeatother

\usepackage{babel}

\makeatother

\usepackage{babel}

\begin{document}

\title{Entanglement Entropy of Gapped Phases and Topological Order in Three dimensions}

\author{Tarun Grover, Ari M. Turner and Ashvin Vishwanath}

\affiliation{Department of Physics, University of California,
Berkeley, CA 94720,
USA}

\begin{abstract}
We discuss entanglement entropy of gapped ground states in different dimensions, obtained on
partitioning space into two regions. For {\em trivial} phases without
topological order, we argue that the entanglement entropy may be obtained
by integrating an `entropy density' over the partition boundary that admits a gradient expansion in the curvature of the boundary. 
This  constrains the expansion of entanglement entropy as a function of system size, and points to an even-odd dependence on dimensionality. For example, in contrast to the familiar result in two dimensions,
a size independent constant contribution to the entanglement entropy can appear for trivial phases in any odd spatial dimension.
We then discuss  phases  with topological entanglement entropy (TEE) that cannot be  obtained by
adding local contributions. We find that in three dimensions there is just one type of TEE, as in two dimensions, that depends linearly on the number of connected components of the boundary
(the `zeroth Betti number'). In $D > 3$ dimensions, new types of TEE appear which depend on the higher Betti numbers of the boundary manifold. We construct generalized toric code
 models that exhibit these TEEs and discuss ways to extract TEE in $D\geq3$.

\end{abstract}

\maketitle

\section{Introduction} \label{sec:intro}
In recent years, surprising connections have emerged between error correction of quantum information and topological condensed matter phases \cite{kitaev2003, nayak2008}.
 At the same time, ideas from quantum information have proved useful in defining topological phases. Two dimensional phases with topological order, such as those
 realized in the context of the Fractional Quantum Hall effect, are gapped phases for which the ground state degeneracy depends on the genus of the space on which
 they are defined. Recent work has shown that they can be identified by the entanglement properties of their ground state wavefunction \cite{hamma2005,levin2006,kitaev2006}.
 The entanglement entropy of a region with a smooth boundary of length $L$ takes the form $S_A=\alpha_1L-b_0 \gamma$, where $\gamma$
 is the topological entanglement entropy, $b_0$ is the number of connected components components
of the boundary of region $A$, and we have dropped the subleading terms. In gapped phases
without topological order, such as band insulators, $\gamma=0$ for a smooth boundary. These predictions have been verified
in the context of a number of specific $D=2$ models \cite{haque, melko, frank, furukawa2007} and in $D=3$ $Z_2$ toric code models \cite{chamon2008}. In this paper, we discuss the general structure of entanglement
entropy for gapped topological and non-topological phases in $D\geq 3$.

One notices that in $D = 2$, the topological entanglement entropy depends only on a topological property of the boundary- in
this case the number of connected components.  There are two equivalent ways of extracting
the topological entanglement entropy \cite{kitaev2006, levin2006}. First,  via the scaling of
the entropy with
 boundary size for smooth boundaries, to extract the constant term. The second, by considering a combination of entanglement entropies of three suitably chosen
 regions $A,\,B, \, C$, so that $-\gamma = S_A+S_B+S_C-S_{AB}-S_{BC}-S_{AC}+S_{ABC}$.

Here, we will discuss analogous questions in $D\geq 3$. In particular, consider a gapped $D=3$ phase, and a region $A$ with a smooth boundary. (1) If the entanglement
entropy $S_A$ contains a constant term, does it necessarily reflect topological order? (2) The boundary of $A$ is a closed two dimensional surface that has two topological
 invariants associated with it - the number of connected components, and the genus (number of handles) of each component. Does this imply there are two distinct types
 of TEEs and correspondingly two varieties of topological order in $D=3$? The answer is {\em no} to both these questions, as we elaborate in this paper. We show that
 even a trivial gapped phase, with no topological order, one can have a constant term in the entanglement entropy in $D=3$ (and in any other odd dimension). Hence, this by
 itself does not signify topological order. Moreover, this constant is generally genus dependent, ruling out a topological origin for a genus dependent entanglement
 entropy. This reduces the number of possible TEE to the same as $D=2$.  We discuss generalization of the Kitaev-Preskill scheme \cite{kitaev2006} to extract the TEE in $D=3$; and why some
 naive extrapolations fail.

 A deeper understanding of TEE is obtained by considering higher dimensions. We show that at least one new topological entanglement entropy appears on going up every two
 dimensions. Thus, while $D=2 \,{\rm and}\, 3$ are similar, a new topological constant does appear in $D=4$ (in $D=2n \,{\rm and} \,2n+1$, there are thus $n$ constants). These
are related to the Betti numbers \cite{footnotebetti} of the boundary. We construct  topological phases that manifest these new TEEs, and explicitly calculate their
 value. These are based on a discrete gauge group $G$. In all dimensions, the TEE
 for discrete gauge theories is $-\log|G|$ per connected surface component, where $|G|$ is the number of elements in $G$. These theories capture both abelian (like the $Z_2$ toric code) and non-abelian phases
and the ground state of these theories correspond to condensate of closed loops. One can also consider more general abelian discrete
gauge theories where the  fluctuating loops are readily generalized to fluctuating $p$-dimensional surfaces. These manifest explicitly in the topological entanglement entropy,
through the appearance of new topological constants that depend on higher Betti numbers. Furthermore, a previously discussed duality between $p$ and $D-p$ theories in $D$
 dimensions \cite{bombin2007, dennis2001, hammawen}, is reflected in the structure of TEE. 

 To isolate topological contributions it is useful to know the structure of entanglement entropy in trivial gapped phases. Since correlations are local in such phases,
we propose an expansion of entanglement entropy $S_A$ based on adding individual contributions from patches on the surface of region A: $S_A=\sum_i S_i$. The entropy
 densities $S_i$ will depend on local properties, such as the local curvature of the surface. One can then expand the entropy density in polynomials of curvature
and its derivatives, similar to the Landau expansion of free energy density \cite{chaikin}. The contribution from higher order terms to $S_A$ are subdominant for large surfaces.
 Interestingly, not every term is allowed in this expansion. When we divide space into a region  $A$ (inside) and $\bar{A}$ (outside), the entanglement entropy of both
 are equal i.e. $S_A = S_{\bar{A}}$. This imposes a Z$_2$ symmetry on the expansion that is {\em unique to ground state entanglement entropy}\cite{corner_note}.
 This has important consequences. Consider for example, $D=2$. The entropy density is not allowed to depend linearly on the boundary curvature $\kappa$, which changes
 sign on interchanging inside and outside. Thus, the expansion of entropy density for a trivial $D=2$ phase is  $S_i=a_0+a_2\kappa^2(r_i)+\dots$, which when integrated around
 the boundary leads to $S_A=\alpha_1 L+ \alpha_3/L+\dots$, where $L$ is the length of the boundary of region $A$. The first term is the area law, and the next term is two orders of $L$ down, due to the Z$_2$ symmetry,
 which eliminates the constant term in total entropy for smooth boundaries. Thus the existence of a constant term in a gapped $D=2$ state implies a {\em non-trivial} phase i.e. topological order. 
In general, this method predicts that for an isotropic, parity invariant state without topological order,
 the entanglement entropy in even (odd) spatial dimensions depends {\em only} on odd (even) powers of $L$, the linear scale of the boundary.
 
 We will assume it is possible to take the continuum limit for all the phases that we consider in this paper. This assumption exclude phases such as a layered $Z_2$ topological ordered phases in $D = 3$ whose
 topological entropy depends on the local geometry of the region $A$.

The paper is organized as follows: in section \ref{sec:decomp} we discuss the general structure of entanglement entropy for gapped phases and explain the basic assumptions underlying our discourse.
In section \ref{sec:EEtrivial} we introduce the aforementioned curvature expansion for entanglement entropy of trivial gapped phases and study its consequences.
In section \ref{sec:EE3d} and \ref{sec:EE4d} we study topological ordered phases in $D = 3$ and $D > 3$ respectively, through extracting the dependence
of entanglement entropy of a region on the topology of its boundary. We also generalize the constructions for extracting topological entropy \cite{kitaev2006, levin2006, chamon2008}. 

\section{Structure of Entanglement Entropy for Gapped Phases} \label{sec:decomp}

In this article, we will assume that the entanglement entropy of a region $A$ can be decomposed into two parts:

\begin{equation}
S_A=S_{A,local}+S_{A,topological}
\label{eq:decomposition}
\end{equation}

\begin{figure}
\centering
\includegraphics[width=.4\textwidth]{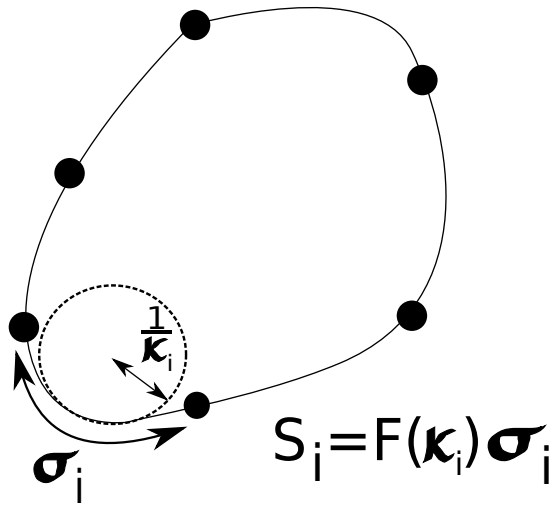}
\caption{The local part of the
entropy of region $A$ is the sum of contributions
of small patches on the boundary.\label{patch}}
\end{figure}

We postpone the underpinnings of this assumption to Sec.\ref{sec:EE3d} when we study topologically ordered phases. Here $S_{local}$ is defined by the property that it can be written as a sum
over contributions from patches located along the boundary of region $A$:

\begin{equation}
S_{A,local}=\sum_i S_i
\label{eq:local}
\end{equation}

where $S_i$ depends only on the shape of the patch $i$, and not on the rest of the
surface or how it fits with other patches, see Fig.\ref{patch},
$\emph{at least}$ if the edge of the patch
connects smoothly to all other patches.

We assume the other contribution
$S_{topological}$ is topologically invariant, i.e., it does not
change as the boundary is deformed unless the topology of the region changes.
If such a term is present and if it cannot be expressed in a local way,
then the phase has long range entanglement, which is the hallmark of topological order \cite{kitaev2003, kitaev2006, levin2006}.

Let us consider the assumptions under which the decomposition (Eqn.\ref{eq:local}) would be possible for a trivial (i.e. not topologically ordered) gapped phase. The reduced density matrix
corresponding to a region $A$ for the ground state wavefunction  may be written as

$$
\rho_A = e^{-H_A}/Z
$$
where $H_A$ is the so called \textit{entanglement Hamiltonian} and $Z = tr(e^{-H_A})$ so that $tr(\rho_A) = 1$. Therefore, we can think of $\rho_A$ as the thermal density matrix at temperature $T = 1$ for the
Hamiltonian $H_A$ and the von Neumann entropy $S_A = -tr(\rho_A\, log{\rho_A})$ as the thermal entropy for this system. Let us define $\tilde{\rho}_A(T) =  e^{-H_A/T}/Z(T)$ where $Z(T) = tr(e^{-H_A/T})$. Clearly,
$\rho_A = \tilde{\rho_A}(T = 1)$ and  $S_A$ obeys the following equation:

\begin{equation}
S_A \equiv S_{A,local} = \int_1^\infty  \frac{dT}{T} \frac{\partial \langle H_A \rangle}{\partial T} \label{integral}
\end{equation}

where $\langle H_A \rangle$ denotes the thermal average of $H_A$ at temperature $T$ with respect to the density matrix $\tilde{\rho}(T)$. We claim that the entanglement entropy
 would admit an expansion such as Eqn.\ref{eq:local}
 if the following conditions are
satisfied.

\begin{itemize}
 \item $H_A$ can be written as a sum of local operators $O$'s i.e. $H_A = \sum_x O(x)$.
  \item There is no phase transition for the Hamiltonian $H_A$ for $T \geq 1$.
\end{itemize}

The first condition along with the fact that all correlations of local operators are short-ranged in a gapped phase  imply that $H_A$ has non-zero support only near the
 boundary of region $A$ (within the distance of correlation length). In other words, the degrees of freedom inside and outside of
region $A$ are coupled only through operators that lie within a distance $\sim \xi$ from the boundary. This implies that $ \langle H_A \rangle = \sum_i h_i $ where $i$
 denotes a point at the boundary of region $A$
and the $h_i$'s depend solely on the properties of the boundary in the vicinity of point $i$.
The second condition implies that the integral in Eqn.\ref{integral} does not admit any singularity so that all terms $S_i$ in the Eqn.\ref{eq:local} are finite.
Physically, this means that the actual system  of
interest is smoothly connected to its $ T = \infty$ zero correlation length system where Eqn.\ref{eq:local} holds trivially.

\section{Entanglement Entropy of trivial gapped phases} \label{sec:EEtrivial}
In this section we will focus on understanding the leading and subleading dependence of $S_{A,local}$ on $L$ for gapped trivial phases of matter. Let us assume that the boundary of region $A$ is smooth, and further
that the phase is isotropic and parity invariant
(consequences of the violation of these assumptions are discussed
at the end of this section and in Appendix \ref{sec:rotpar}).
Then, as we will see below, in the absence of topological order, only alternate terms in the power series expansion of $S_A(L)$ appear:

\begin{equation}
S_{A,local}(L) = \alpha_1 L^{D-1}+\alpha_3 L^{D-3}+\alpha_5 L^{D-5}+\dots \label{expanL}
\end{equation}

i.e. only those with odd co-dimension exponent can appear. This expansion implies a distinction between even and odd dimensions:
in even dimensions, any constant contribution to the entanglement
entropy must come from $S_{A,topological}$, and thus
indicates topological order. In odd spatial dimensions, a constant term may appear in the local entropy,
making it more difficult, though still possible,
to isolate topological contributions (note that these conclusions
apply only to smooth boundaries in rotationally symmetric
systems; a corner can produce a constant term
for even dimensions as well as odd ones).

Let us consider some instances of Eqn.\ref{expanL} that will
motivate its derivation. First, it is well known that in $D=2$,
a constant term in $S_A(L)$ implies the presence of topological order \cite{kitaev2006, levin2006}.
On the other hand, in $D=3$, a constant term can appear for a non-topologically ordered phase, such as a gapped scalar field.
Consider for a moment \cite{casini2009} a {\em massless} field,
where it is known that the entanglement scales as $S_A \sim L^2 + log(L)$
for a spherical ball of radius $L$. Now, providing a mass $m$ to the scalar
field would cut off the $log(L)$ term and instead lead to a constant contribution proportional to $log(1/m)$. We verified this explicitly using a numerical calculation similar to Ref. \cite{srednicki}.
Interestingly, when the surface of region $A$ is flat,
then there is no constant contribution as we show in Appendix A (a flat boundary is possible when the total system has the  topology of a three torus $T^3$; then $A$ may be taken
 to have the geometry $T^2 \times l$, where $l$ is a line segment).
This indicates that the presence or absence of a constant term may have something to do with the curvature of the boundary of region $A$. In the next subsection, we make this statement
precise and explain the observations made above.

\subsection{Entropy Density Functional} \label{sec:local}

\begin{figure}
\centering
\includegraphics[width=.5\textwidth]{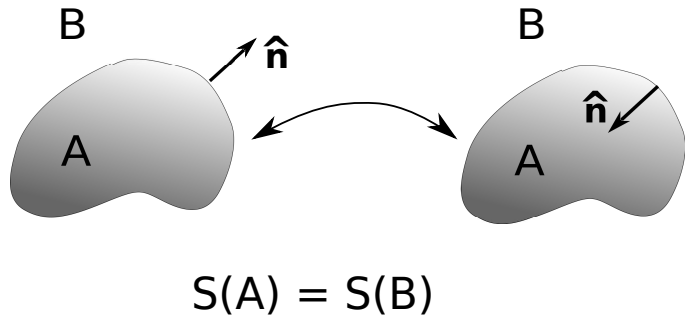}
\caption{Illustration of the Z$_2$ symmetry for the curvature expansion discussed in the text.\label{fig:inout}}
\end{figure}

\begin{table*}
\begin{tabular}{|l|c|c|p{4cm}|}
\hline
&  \textbf{Full symmetry} & \textbf{All symmetries broken} & \textbf{Broken parity alone}\\
\hline
$\mathbf{D=2}$ & All even terms $L^0,L^{-2}$ & Constant term & All even terms\\ \hline
$\mathbf{D\geq 3}$ \textbf{odd}& All Odd terms $L^{D-2}, L^{D-4},\dots$ & Nothing forbidden &
Odd terms with positive exponents and also $L^{-1}$ if $D\equiv
3(\mathrm{mod\ 4})$\\ \hline
$\mathbf{D\geq 4}$ \textbf{even} & Even terms $L^{D-2},L^{D-4},\dots$ &Nothing forbidden &
Even terms\\
\hline
\end{tabular}
\caption{\label{table:mahogany}Terms in the entropy forbidden by
symmetries.
The three columns describe systems with rotational and parity
symmetry, no spatial symmetry, and rotational but not parity symmetry. The
entries list the scaling of terms $L^k$ that are forbidden
from appearing in the entropy.}
\end{table*}

Let us consider a region in two dimensions for concreteness.
We postulate that the local entropy $S_{A,local}$ is given by the following integral:

\begin{equation}
S_A = \sum_i S_i = \int d\sigma F(\kappa,\partial\kappa,\dots)
\label{eq:continuous}
\end{equation}

where $F(\kappa,\partial\kappa,\dots)$ is the ``entropy density functional".
In a gapped phase, the entropy $S_i$ of a patch larger than
the correlation length
can  depend only on the properties
of the patch, such as its length $\Delta\sigma_i$ and
curvature $\kappa_i$, as well as derivatives of the latter,
and must be proportional to $\Delta\sigma_i$.
Hence $S_i=\Delta\sigma_i F(\kappa_i,\partial^n\kappa_i)$. Taking
the limit where the patches become microscopic
compared to $L$ (but greater than $\xi$)
leads to Eq. (\ref{eq:continuous}).

The entropy density functional always satisfies a $Z_2$ symmetry, which
is the key to understanding the $L$-dependence of the entropy.
The symmetry results from the fact that, if $A$ and $B$
are complementary regions, then $S_A=S_B$.
Therefore, changing `inside' to `outside' keeps the entanglement entropy
invariant. Now, under this transformation, radii of curvature clearly
change sign $\kappa \rightarrow -\kappa$, and this constrains
the entropy density functional (Fig.\ref{fig:inout}). As an illustration of this $Z_2$ symmetry, consider the form of the functional $F$ for a gapped two dimensional system.
 On a smooth boundary, one can expand the function $F$ in a Taylor series, retaining the first few terms:

\be
F(\kappa_i,\partial^n_\sigma\kappa_i)=a_0+a_1\kappa_i + a_2\kappa_i^2+b_2\partial_\sigma\kappa_i+\dots \label{expan2d}
\ee

The first term gives the boundary law $S_A=a_0 L$. The second term, if it is present, would give a constant contribution $\oint d\sigma\kappa = 2\pi$
for the curve shown, which would be
 a non-universal constant contribution.
However, such a term is in fact forbidden by the $Z_2$ symmetry, since
the term is odd in $\kappa$.
The term $\kappa^2$ gives the next contribution to the entropy
that is proportional to $\oint d\sigma\kappa^2$. If the shape of region $A$ is kept fixed, then this contribution scales as $1/L$. 
The term  $\partial_\sigma\kappa$ is allowed by the $Z_2$ symmetry
(since both the derivative and the radius of curvature change sign,
assuming that the direction of the curve is set by a `right hand rule', whereby the arc length increases along a specific direction), yet it still
vanishes because it is a total derivative. Generalizing these arguments,
one finds that $S_A=\sum_{k=0}^\infty \alpha_{2k+1}L^{1-2k}$.

In general dimensions, we find similar results if we continue
to assume rotational, parity, and translational symmetry. These
assumptions imply that $F$ can depend only on the metric
tensor $g^{\alpha\beta}$ and
on the extrinsic curvature
of the surface.  The latter tensor does not appear when considering intrinsic properties of a manifold (as in general relativity). However,
 entanglement entropy
does depend on the embedding of the boundary $\partial A$, since it is measuring
the entanglement of the degrees of freedom in the space around the surface.
The extrinsic curvature is a tensor $\kappa_{\alpha\beta}$ with two indices
(see Appendix \ref{sec:primer} for a short primer on the requisite differential
geometry).
Thus each term in $F$ contains some number of factors of $\kappa_{\alpha\beta}$
and its covariant derivatives, with all the indices contracted by factors
of $g^{\gamma\delta}$ (if parity is broken, the antisymmetric volume tensor
$\gamma^{\alpha_1\dots\alpha_{D-1}}$ is allowed as well, Appendix \ref{sec:rotpar}).

The inside/outside symmetry further limits the form
of the terms in $F$: it implies
that each term in $F$ includes an even number of factors $n_{\kappa}$ of
$\kappa$. The total order of all the derivatives $n_D$
must also be even. This follows from rotational symmetry. For rotational symmetry to be respected, one has to
contract all the lower indices with the tensor $g^{\gamma\delta}$.
This leads to an even number of lower
indices, that include the derivatives as well as the curvature indices. Since the curvature tensor is of even rank, the number of derivatives
$n_D$ has to be even. Putting everything together, one finds that the contribution to the entropy density $F$ scales as $L^{-(n_\kappa+n_D)}$ that clearly has an even exponent,
explaining why only alternate terms appear in the entropy, Eq. \ref{expanL}.

When rotational or inversion symmetries are broken spontaneously or by applying a field, additional terms appear in
the entropy, as summarized in
Table \ref{table:mahogany}. We provide the details leading to these results in Appendix \ref{sec:rotpar}.

The local entropy can also contain topology-dependent terms e.g. the term $\int G dA=4\pi \chi$
in three dimensions where $G$ is the Gaussian curvature which is the determinant of the  matrix $\kappa_{\alpha \beta}$.  Note that this term is compatible with the symmetry
$\kappa\rightarrow -\kappa$, since $G$ is quadratic in $\kappa$. Hence,
as mentioned earlier, the presence of a term in the entropy that is proportional to the Euler characteristic does
not necessarily correspond to topological order.
In general dimensions, $\int \mathrm{det}\kappa dA$ is topological,
but it is only symmetric in odd dimensions, where it
is proportional to the Euler characteristic of the boundary
(in general, it is proportional to the Euler characteristic
of the region itself \cite{gauss-bonnet}).

\section{Topological Entanglement Entropy in $D= 3$} \label{sec:EE3d}

We now turn to the topological part of the entanglement entropy. Our starting point is  Eqn. \ref{eq:decomposition} which we rewrite here for convenience:
\begin{equation}
S_A=S_{A,local}+S_{A,topological}
\label{eq:decomposition2}
\end{equation}
This decomposition is what enables the extraction of topological entropy using Kitaev-Preskill \cite{kitaev2006} or Levin-Wen \cite{levin2006} constructions for two-dimensional
topological ordered phases. The assumptions underlying this equation are somewhat tricky. Though Eqn.\ref{eq:decomposition2} holds for toric code models in all dimensions and there is strong numerical evidence
 that it also holds for
many interesting two dimensional topological ordered states such as $Z_2$ spin liquids, quantum dimer models and various quantum Hall states  \cite{furukawa2007, haque, melko, sierra2010, frank},
Eqn.7 surreptitiously rules out
a layered $Z_2$ topologically ordered state.  Such a state would lead to a correction in entanglement entropy $\Delta S_A = -\gamma_{2D} L_z$,
for layering perpendicular to the $z$ direction. Here $\gamma_{2D}$ is the topological entanglement entropy associated with the theory living in each layer. Clearly, $\Delta S_A$ is not
topologically invariant. The assumptions underlying Eqn. \ref{eq:decomposition2}
 most likely also do not apply to the self-correcting code state of Ref. \cite{haah2011}; this state (whose
ground state degeneracy depends on the divisibility of system size by powers of $2$,
for example) illustrates why we need to make an assumption of this type. Nevertheless, we will briefly discuss the topological entanglement entropy of layered $Z_2$ state in Appendix \ref{sec:layer}.

\textit{Independent contributions to $S_{topological}$:} The boundary $\partial A$ of a three dimensional region $A$ is a compact manifold that is characterized by Betti numbers $b_0$ and $b_1$ (note that for
compact manifolds $b_2 = b_0$). As we show in Appendix \ref{sec:linear} in three dimensions $S_{A,topological}$ is a linear function of $b_0, b_1$,
say, $S_{topological} = -\gamma_0b_0-\gamma_1b_1$ (we assume that the space in which region $A$ is embedded has the topology of $\mathbb{R}^3$, otherwise more complicated
dependence is possible in principle). This might lead one to suspect that there are two different kinds of topological orders in three dimensions, namely,
those corresponding to a non-zero $\gamma_0$ and $\gamma_1$ respectively. However, $b_0, b_1$ are related to the Euler characteristic $\chi$ through
$2b_0 - b_1 = \chi$. Thus, one may redefine $S_{A,local}'=S_{A,local}+\alpha\chi$ and $S_{A,topological}'=S_{A,topological}-\alpha\chi$ without
changing the entropy \textit{and} $\alpha$ may be adjusted so that the $b_1$ dependence of $S_{topological}$ is canceled out.
Here the term $\alpha\chi$ may be thought of as both local and topological. It is local, because the Euler's formula, $\chi=V-E+F$ gives a local expression
for this term, where $V$, $E$, and $F$ are the number of vertices, edges, and faces into which $\partial A$ is divided (alternatively,
in a continuum theory, $\alpha\chi$ can be incorporated into the entropy density $F$ since $\chi$ is the integral of the Gaussian curvature). It is also
topological, because $\alpha\chi$ is independent of how the surface is divided up into regions. The upshot of this discussion is that there is only one kind of topological entropy in three dimensions.

\textit{$Z_2$ string and $Z_2$ membrane models:} As an alternative way to understand the above result, let us study specific models whose
topological entanglement potentially depends on different Betti numbers. Consider the following model of $Z_2$ gauge theories consisting of spin-$1/2$ degrees of freedom that live
on the links of a three-dimensional cubic lattice:

\be
H_{string} =  -\sum_{\Box} \,\,\prod_{l\in \Box} \tau_{z,l}  - h \sum_l \tau_{x,l} \label{hstr3d}
\ee

where $\Box$ denotes a plaquette of the cubic lattice, the operators $\tau_{x,l}, \tau_{z,l}$ live on the links $l$ of the lattice. The above Hamiltonian is supplemented with the constraint (`Gauss law')
$ \prod_{l \in \,vertex} \tau_{x,l} = 1 $ to impose the absence of $Z_2$ charges in the theory. Because of this constraint the gauge invariant degrees of freedom in this model consist of closed loops
$\mathcal{C}$ on the edges of the lattice. In the deconfined phase of the gauge theory, $ |h| \ll 1$, the loops condense
because they do not cost much energy. The entanglement entropy of this model for a region $A$ depends only on the Betti number $b_0$ of $\partial A$, since each component of the boundary places a separate 
constraint on the loops
that intersect the boundary $\partial A$. Let us take Kitaev's `toric code limit' of the above model \cite{kitaev2003} by setting $h=0$.
In this limit, the constraint commutes with the Hamiltonian and can be included as a part of it. Hence
the model may be written as

\bea
H_{string,h=0} & = & -\sum_{\Box} \,\,\prod_{l\in \Box} \tau_{z,l}  - \sum_{vertex}\,\, \prod_{vertex \in \,l} \tau_{x,l} \nonumber \\  \label{hstr3d2}
\eea

Interestingly, the ground state of Eqn.\ref{hstr3d2} may be reinterpreted as a superposition of closed \textit{membranes}. This is seen as follows.
The first term in the Hamiltonian may be regarded as the constraint $\prod_{l\in \Box} \tau_{z,l} = 1$. Now consider the dual lattice each of whose plaquettes 
is pierced by a link `$l$' of the original cubic lattice. A surface can be defined by the plaquettes of the dual lattice pierced by  $\tau_{z,l}=-1$ bonds. Due to the constraint, this surface is closed.
  Thus there is no distinction between condensed loops and condensed membranes in this case\cite{bombin2007,hammawen},
consistent with the fact that in three dimensions there is only one kind of topological entanglement entropy.

\textit{Discrete gauge theories in $D=3$:} Before moving on to the discussion of topological entropy in general dimensions, let us derive the entanglement entropy corresponding to a discrete gauge theory with
 general gauge group $G$ for
Kitaev model \cite{kitaev2003} on a cubic lattice \cite{thanklevin}

\be
H = - t \sum_{p} \delta(g_1 g_2 g_3 g_4 = e) - V \sum_{s, g} L^{1}_g L^{2}_g L^{3}_g L^{4}_g
\ee

Here `p' stands for a plaquette, `s' for a star (i.e. six links emanating from a vertex) while $g$'s are the elements of group $G$ with size of
group being $|G|$. For non-abelian groups one needs to chose an orientation of the links so that for opposite orientations, the group element on a link
is $g$ and $g^{-1}$. The operators $L^{g}$ live on the links and and their action is described by $L_{g_1} |g_2 \rangle = |g_1 g_2 \rangle$ or
$L_{g_1} |g_2 \rangle = |g_2 g_1^{-1} \rangle$ depending on whether $g_1$ points away from or towards the
vertex at which the action of $L_g$ is being considered. The ground state of $|\Phi\rangle$ of $H$ is given by

\be
|\Phi\rangle  = \sum_{\{g\},g_1 g_2 g_3 g_4 = e \,\forall \,plaquettes} |\{g\}\rangle
\ee

Let us divide the entire system into region $A$ and $B$ and assume that the boundary is made up of plaquettes of the lattice. The links along the boundary are labeled by the group elements $h_1, h_2, ...,h_n$. 
The Schmidt decomposition of $|\Phi\rangle$ reads

\be
|\Phi\rangle = \sum_{\{h\}} |\phi\rangle^{\{h\}}_{in} \otimes |\phi\rangle^{\{h\}}_{out}
\ee

where

\be
|\phi\rangle^{\{h\}}_{in}  = \sum_{\substack{\{g\},g_1 g_2 g_3 g_4 = e \,\forall \,plaquettes \in A,\\  g_i = h_i \,\textrm{for}\, i \in \partial A}} |\{g\}\rangle
\ee

and $|\phi\rangle^{\{h\}}_{out}$ is defined similarly. All the states in the Schmidt decomposition enter with the same weight and are orthogonal, therefore the entanglement entropy is
the logarithm of the number of states. These may be counted by finding all the configurations for $\{h\}$ that satisfy the following constraint:  the product of the $\{h\}$'s around
any closed loop on the boundary must equal the identity \cite{hamma2005, levin2006}s.  This includes contractible as well as noncontractible loops on the surface, and each independent loop reduces the total
number of configurations by a factor of $|G|$, leading to

\bea
S & = & log (|G|^{V-1}) \nonumber \\
& = & V log(|G|) - \gamma \label{eq:discretetee}
\eea

where $V$ is the number of vertices on the boundary
and $\gamma = log(|G|)$ is the topological entanglement entropy. This result for topological entanglement entropy is identical to that for discrete gauge theories in $D=2$.

\subsection{Extracting Topological Entanglement Entropy in $D =3$}

\begin{figure}
 \includegraphics[width=240pt, height=220pt]{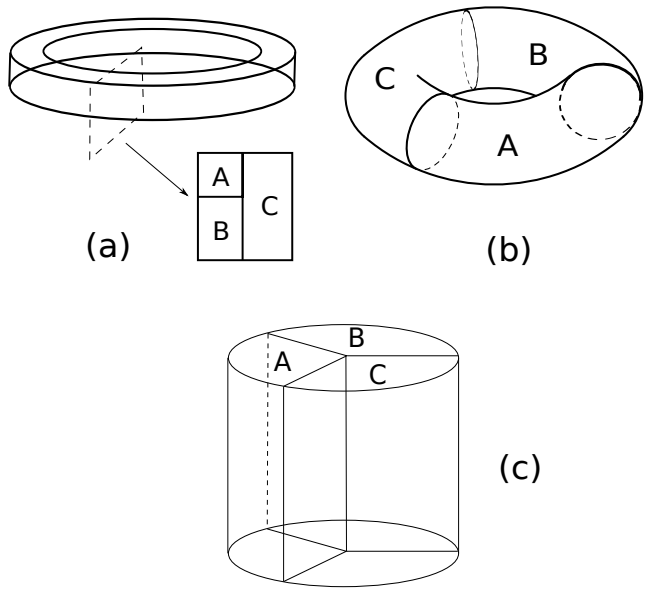}
\caption{Fig.(a) and (b) show two valid $ABC$ constructions (Eq. \ref{kitpres}) in three dimensions that can be used to extract the topological entanglement entropy. In Fig.(a) the cross-section of a torus has
 been divided into three tori $A, B$
and $C$ while in Fig.(b) a torus that has been divided into three cylinders $A, B$ and $C$. The Fig.(c) shows an \textit{invalid} construction as explained in the text. In all three figures,
we define region $D$ to be the rest of the system.}
\label{fig:3dtee}
\end{figure}

In the spirit of Ref. \cite{kitaev2006, levin2006}, we would like to combine the total entanglement entropies of
certain regions in such a way that the local part of the entropy cancels out while topological part survives.

The Kitaev-Preskill construction, which succeeds in
this task in two dimensions can be modified so that it works
for three dimensions as well (Ref.\cite{chamon2008} describes an extension of the Levin-Wen scheme to $D=3$ ).
The construction involves three regions $A,B,C$ embedded
inside region $D$:

\be
-\gamma_{topo} = S_A + S_B + S_C - S_{AB} - S_{BC} - S_{CA} + S_{ABC} \label{kitpres}
\ee

In two dimensions, the regions are taken to be three $120^\circ$ segments
of a circle. In three dimensions, a direct generalization of the two-dimensional
construction (dividing a \emph{cylinder} into three sectors as in Fig.\ref{fig:3dtee}c) fails to be topologically invariant
because the changes in the entropy
near the points at the top and bottom of
the cylinder where $A,B,C$ and $D$ all meet do not cancel.
However, the regions such as
the two shown in Fig.\ref{fig:3dtee}a and 3b can be used where such points do not exist.
For example, if one deforms the circle at which regions $A,B,D$ all meet
(in either geometry), then

\bea
\Delta \gamma_{topo} & = & -[\Delta (S_A - S_{CA}) + \Delta (S_B - S_{BC}) + \Delta (S_{ABC} - S_{AB})] \nonumber \\
& = & 0
\eea

The last equation follows because each of the three terms in the brackets could be thought of as the difference between entropies of two regions that differ by addition of region $C$ (that
is located far from the point where $A, B$ and $D$ meet). Since each
region has a single boundary component, $\gamma_{topo}=S_{topological}$.
In Appendix \ref{sec:valid3d}, we detail the general requirements for a construction that would always yield a topological invariant.

Based on our earlier discussion of curvature expansion for entanglement entropy, we note that in the special case of a completely flat boundary between a region
and the rest of the system, the constant term in the entanglement entropy corresponding to that region can indeed be identified with topological entanglement entropy \cite{footnote_degen}. This can
be realized by taking the total system to be $T^3$ and region $A$ as $T^2 \times l$ where $l$ is a line segment (similar to the calculation of
entanglement entropy for a  free scalar in Appendix \ref{sec:scalar}).

\section{Topological Entanglement Entropy in $D > 3$} \label{sec:EE4d}

\textit{Independent terms in $S_{topological}$ in arbitrary dimensions:} Following our discussion of topological entanglement entropy in $D = 3$, in this section we study the independent contributions to
$S_{topological}$ in a general dimension $D>3$. The boundary $\partial A$ of a $D$-dimensional region $A$ is a compact manifold that is characterized by Betti numbers, $b_0,\dots,b_{D-1}$ that
describe various orders of connectivity of the surface (see e.g. \cite{nakahara}).

We will assume a linear relationship,
$S_A=-\sum_{k=0}^{D-1} \gamma_k b_k$. In principle, in higher dimensions the entanglement entropy could depend on more subtle
topological properties of the boundary, but we will  focus only 
on this form.
Further, as we will see below, this form turns out to be sufficient for Kitaev models that describe
discrete $p$-form gauge theories ($p \ge 1$) in arbitrary dimensions.

To see how many types of topological entropy can exist in higher
dimensions, first  note that
for compact manifolds, the Betti numbers have a symmetry,
$b_k=b_{D-1-k}$ and hence the sum may be cut short, at
$k=\lfloor\frac{D-1}{2}\rfloor$.
Furthermore, owing to the relation $\chi=\sum_{k=0}^{D-1} (-1)^k b_k$, in all odd dimensions a
part of the topological entropy may be absorbed into the local entropy,
reducing the number of coefficients by one more.
Hence there are $n$ topologically nontrivial
contributions to the entanglement entropy in $2n$ and $2n+1$
dimensions:
\begin{equation}
S_{A,topological}=\begin{cases}
-\gamma_0 b_0 - \gamma_1 b_1 - \dots - \gamma_{\frac{D}{2}-1}b_{\frac D2-1}, & \text{if }D\text{ is even}\\
-\gamma_0 b_0 -\gamma_1 b_1+\dots - \gamma_{\frac{D-3}{2}}b_{\frac {D-3}{2}}, & \text{if }D\text{ is odd}\end{cases}.
\label{gamma_d}
\end{equation}

Precisely such a hierarchy of states associated
with different Betti numbers has been arrived
at by Ref. \cite{bombin2007}  by
constructing a sequence of
Kitaev `toric-code' type models where the ground
state is a superposition of all $p$-dimensional manifolds on a lattice (for $1\leq p\leq D-1$).
This state is dual to the superposition of all $q=D-p$ dimensional manifolds,
so the number of distinct models is $\lfloor\frac{D}{2}\rfloor$, the same
as the number of types of topological entropies.

\textit{$S_{topological}$ for gauge theories in arbitrary dimensions:} Similar to three dimensions, one may study models of discrete gauge theories to understand these results.
For example, on a hypercubic lattice in $D=4$, the string and membrane theories describe very different ground states \cite{dennis2001} and unlike $D = 3$, the membrane
theory is now dual to itself, not to the string phase. Explicitly, in the `toric code limit' \cite{kitaev2003, dennis2001} these two theories are given by 

\be
H_{string} =  -\sum_{\Box} \,\,\prod_{l\in \Box} \tau_{z,l}  -  \sum_{vertices} \, \prod_{vertex \in \, l} \tau_{x,l} \label{hstr3dkit}
\ee

\be
H_{membrane} =  -\sum_{l} \,\,\prod_{l \in \Box} \sigma_{z,\Box}  - \sum_{cubes} \, \prod_{\Box \in \,cube} \sigma_{x, \Box}  \label{hmem3dkit}
\ee

As we show now, the entanglement entropy of the model in Eqn. \ref{hstr3dkit} in four dimensions
depends on the Betti number $b_0$ of $\partial A$ while that corresponding to model in Eqn. \ref{hmem3dkit} depends on the difference
$b_1 - b_0$. For the sake of generality, let us derive the entanglement entropy of a generalized toric model in arbitrary spatial dimensions $D$ whose ground state is given by sum over all closed
$d_g$ dimensional membranes. This ground state describes deconfined phase of a $d_g$-form abelian gauge theory.
These membranes intersect the boundary $\partial A$ of region $A$ in closed membranes of dimension $d_g -1$, \textit{with the restriction that these
intersections are always boundaries of a membrane of dimension $d_g$ contained in $\partial A$}. For example,
consider the entanglement of membrane model in Eqn.\ref{hmem3dkit} in $D = 3$  when the boundary of region $A$ is a torus $T^2$ (note that the form of Hamiltonian for membrane theory is identical
in $D = 3$ and $D = 4$).
When a closed membrane intersects $\partial A = T^2$, one sees that one can only obtain an even number of closed loops along any non-contractible cycle of $T^2$, which would therefore form
the boundary of two dimensional membrane. Returning to the general case,
let us denote the number of independent $n$-dimensional membranes that belong to $\partial A$ by $C_n$ and those that are boundary of a $n+1$-dimensional membrane by $B_n$.

Using the definition of Betti numbers \cite{nakahara} and simple linear algebra, one finds that the entanglement entropy $S_A$

\be
S_A \propto  \sum_{n=0}^{d_g-1} (-)^{d_g-1 + n} C_n - \sum_{n=0}^{d_g-1} (-)^{d_g-1 + n} b_n
\ee

Since the $C_n$ are expressed in terms of local quantities such as the number of edges, vertices etc. that lie on the boundary without any additional constraint,
we identify the first sum as $S_{local}$ and the second as $S_{topological}$. The proportionality constant depends on the
gauge group and akin to three dimensions equals $log(|G|)$ where $|G|$ is the number of elements in the abelian gauge group (note that the calculation of TEE in $D$ = 3 (Eqn. \ref{eq:discretetee})
applies to abelian as well as non-abelian discrete gauge theories). Therefore

\be
S_{topological} = - log(|G|) \sum_{n=0}^{d_g-1} (-)^{d_g-1 + n} b_n
\ee

\emph{Extracting topological entropy in Four Dimensions}  We will restrict our discussion of extracting $S_{topological}$ to four dimensions.
For a given region  $A$, from Eqn. \ref{gamma_d} one has
$S_A = S_{A,local} - b_0 \gamma_0 -b_1 \gamma_1$ and one would like to have a
construction similar to Levin-Wen \cite{levin2006} and/or Kitaev-Preskill \cite{kitaev2006} that enables one
to extract the topological numbers $\gamma_0$ and $\gamma_1$.

We extract $\gamma_1$ by a generalization
of the  construction in \cite{levin2006}.  Let region $A$ have the topology of $B^2\times S^2$. Region $B$ is $A$ with a channel cut in it and has topology of  $B^4$. Finally,
region C has a second identical channel cut out opposite to the first one and has topology $S^1\times B^3$

Now,
$S_A-2S_B+S_C$ is topologically invariant just as in two
dimensions
and the Betti numbers of the bounding surfaces are:

\bea
b_0(\partial A) & = & 1, \,\,\,\,\,b_1(\partial A) = 1 \nonumber \\
b_0(\partial B) &= & 1, \,\,\,\,\, b_1(\partial B) = 0 \nonumber \\
b_0(\partial C) &= & 1, \,\,\,\,\, b_1(\partial C) = 1 \nonumber
\eea
Hence $(S_A - S_B) - (S_B - S_C) =- 2 \gamma_1$. Since $\gamma_1 \ne 0$ for the membrane Kitaev model $H_{membrane}$ (Eq.\ref{hmem3dkit}) while it is zero for the string model $H_{string}$ (Eq.\ref{hstr3dkit}) in $d = 4$,
this construction measures membrane correlations.

To isolate $\gamma_0$,
the analogous procedure, but with $A$ being $B^3\times S^1$ suffices.
The combination $(S_A-S_B)-(S_B-S_C)$ gives $\gamma_0+\gamma_1$, and
this may be combined with the previous construction to extract
both $\gamma_0$ and $\gamma_1$. This construction selectively measures string correlations since $\gamma_0 + \gamma_1$ is zero for $H_{membrane}$.

The Kitaev-Preskill construction of dividing a disc into three triangles that meet at the center is readily extended to any even dimension. In $D=4$ consider dividing the ball $B^4$ into five `pentahedra' 
that meet at the center. The combination

\be
\Delta(\{S\}) = \sum_i S_i - \sum_{i<j} S_{ij} + ... + S_{12345}
\ee

is topologically invariant and gives $-\gamma_0$.
Here $S_{i_1 i_2...i_n}$ denotes the entanglement entropy corresponding to the region $A_{i_1}  \cup A_{i_2} ... \cup A_{i_n}$.

\section{Discussion and Conclusion}

In this paper, we discussed the qualitative structure of the entanglement entropy for gapped phases.
We introduced the concept of `entanglement entropy density' whose integral over the boundary of a region $A$ yields the
entanglement entropy of region $A$. For gapped trivial phases the symmetry constraints on the entropy density, including the inside-outside exchange symmetry $S_A = S_{\overline{A}}$, naturally lead to the leading
 and subleading dependence of the entanglement entropy on the linear size of a given region.

In the second half of the paper, we studied the topological entanglement entropy $S_{topological}$ of topologically ordered systems in various dimensions. A key result was that in $D=3$ there
 is a single category of TEE, as in $D=2$, that depends linearly on the number of connected components of the boundary. This constrains the possible forms of topological order in $D=3$.

We briefly discussed TEE in higher dimensions - using generalized Kitaev toric code like models (i.e. deconfined
phases of $p$-form discrete gauge theories) to realize various topologically ordered phases. In $D=4$ we find two categories of TEE . In general, one new category of TEE appears each time
 the dimension is raised by two. This even-odd effect is understood as follows. $S_{topological}$ depends on the Betti numbers of the boundary of region $A$. In odd spatial dimensions,
 the Gauss-Bonnet theorem relates Betti numbers to the curvature of the boundary of region $A$. This implies that there is one linear combination of Betti numbers that can be
 expressed as an integral of a local property of the boundary (such as curvature), and is thus not an independent topological contribution to the entanglement entropy.  We also mentioned how
to extract $S_{topological}$  by a generalization of the $D=2$ Kitaev-Preskill and Levin-Wen constructions.

Potentially, in $D\geq 4$, $S_{topological}$ may depend not only on Betti numbers of the boundary manifold, but on more subtle topological properties such as its homotopy group.
 If such phases do exist, then entanglement entropy could shed light on the classification of manifolds. Lattice 3D models realize a richer variety of topological phases than the
 isotropic phases considered here. For example, there exist  layered $Z_2$ topologically ordered phases, which retain a two dimensional character despite coupling between layers.
 Another example is the self correcting quantum memory of Ref.\cite{haah2011}. For these, the separation between the topological and the local part of the entanglement entropy is
 not obvious. General statements about entanglement in such topological phases remain for future work.

One might also consider a curvature expansion for the fluctuations of a conserved quantity such as particle number or total spin, inside a region $A$.
Intuitively, these would be a property of the boundary of region $A$ \cite{klich2006, klich2006_2, lehur2010,levitov2009}. Indeed, akin to entanglement entropy, one has $F_A = F_{\overline A}$ where
$F_A = \sqrt{\langle ( \sum_{r \in A} O_r )^2 \rangle - \langle\sum_{r \in A} O_r \rangle^2} $ is the variance of $O$ inside the region $A$. Therefore,
a curvature expansion for $F_A$ would inherit many of the arguments we used to derive the leading and sub-leading behavior of the quantity $F_A$, and can provide a framework to understand known
 results  \cite{lehur2010,klich2006, klich2006_2, levitov2009, helling2011, swingle2010}.

Finally, it may be possible to learn more about the systematics of the size dependence of entanglement entropy in {\em gapless} phases by a generalization of the curvature expansion
 under certain conditions.  Many gapless systems such as massless scalar/Dirac fermion also follow an area law and have an expression for entropy with interesting parallels to Eqn.\ref{expanL}.

\textit{Acknowledgements:} We thank Michael Levin for illuminating discussions and Matt Hastings for helpful comments on the manuscript. Support from NSF DMR- 0645691 is acknowledged.

\appendix

\section{Absence of constant term for massive scalar in the absence of curvature} \label{sec:scalar}

We are interested in the entanglement entropy of a massive scalar field in three dimensions when the region $A$ has a geometry
 $T^2 \times l$ where the torus $T^2$ extends along the directions $1,2$ while $l$ is a line segment of length $l$ along the direction $3$.
In particular, we want to show that the constant part of the entanglement entropy is in fact exactly zero i.e. $S = A \,L^2 + O(1/L)$.
This is consistent with the curvature expansion of entanglement entropy (Eqn. \ref{eq:continuous}) since now the region $A$ does not  have any extrinsic curvature.

The Euclidean action $\mathcal{S}$ is given by

\be
\mathcal{S} = \int |\phi(k_1,k_2,k_3,\omega)|^2 (m^2 + \omega^2 + \gamma^2(3-cos(k_1) - cos(k_2) - cos(k_3)))
\ee

where we impose periodic boundary conditions in all directions and we have set the lattice spacing to unity.
Periodic boundary conditions imply momenta $k_1, k_2$ remain good quantum numbers even after making the partition. The total entanglement entropy may therefore be written as

\be
S = \sum_{k_1,k_2} S_{1D}(M(k_1,k_2,m))
\ee

where $S_{1D}(M)$ is the entanglement entropy of a one dimensional massive scalar theory with mass $M =  \sqrt{m^2 + \gamma^2(2-cos(k_1) - cos(k_2))}$. Using  \cite{calabrese}
$S_{1D}(M) \propto -log(M^2)$,

\be
S \propto -\sum_{k1,k2} log(m^2 + \gamma^2(2 - cos(k_1) - cos(k_2)))
\ee

Using Euler-Maclaurin formula, one finds the following expression for $S$, correct to $O(L^0)$:

\bea
S & \propto & I_1 + I_2
\eea

where

\bea
-I_1 \simeq L^2 log(m^2) + L^2 \int_{0}^1 dt\, log(1 + \gamma^2/m^2(1-cos(2 \pi t))) \nonumber
\eea

and

\bea
& & -I_2 \simeq \nonumber \\
& & 2 \pi \gamma^2 L^2 \int_{u = 1/L}^1 \int_{t = 0}^1 \frac{t sin(2\pi t)}{m^2 + \gamma^2(2 - cos(2 \pi t) - cos(2 \pi u))} \nonumber \\
& & + 2 \pi \gamma^2 L \int_{0}^1 \frac{t sin(2 \pi t)}{m^2 + \gamma^2(1-cos(2 \pi t))} \nonumber \\
& & \simeq  2 \pi \gamma^2 L^2 \int_{u = 0}^1 \int_{t = 0}^1 \frac{t sin(2\pi t)}{m^2 + \gamma^2(2 - cos(2 \pi t) - cos(2 \pi u))} \nonumber
\eea

Clearly neither $I_1$ nor $I_2$ contribute to a constant term in the entanglement entropy. Therefore, as anticipated from the curvature expansion, the entanglement entropy
$S = I_1 + I_2$ does not contain a constant term for the $T^2 \times l$ geometry and is proportional to $L^2$  upto $O(L^0)$.

\section{Additional terms in Entanglement entropy in the absence of rotational/parity symmetry} \label{sec:rotpar}

\subsection{Broken Rotational Symmetry} \label{sec:rot}
When rotational symmetry is broken, all powers of $L$ (after
the area-law term) are present, except in two dimensions.   In
$2$-dimensions, there
is no constant term when the boundary
of the region is smooth.  So if rotational
symmetry is broken, it is not as easy to recognize a topological phase in
even dimensions higher than two without resorting to the $ABC$ construction (Eqn. \ref{kitpres}) or its analogs.

In one dimension, when symmetry is broken, it is convenient
to express the entropy in terms of $x(s)$ and $y(s)$, the parametric
equation for the boundary.  Because rotational symmetry is broken,
there is no requirement that these terms appear symmetrically.
However, because of translational symmetry,
the entropy is a function only of the derivatives of these
functions.   The only requirement is that
the entropy density must be symmetric under $s\rightarrow -s$.
Otherwise the entropy depends
on whether s is measured clockwise or counterclockwise
around the region, and this violates the symmetry between
the inside and outside.
(Clockwise is defined relative to a choice of the region's inside, as is
familiar from using residues to evaluate integrals in the
complex plane.)
The expression $\int (\frac{d^2x}{ds^2})^3 ds$ is symmetric
and it scales as $\frac{1}{L^2}$. One can check that the
integral is nonzero for an ellipse (assume the ellipse has
a small eccentricity so that the integral can be evaluated--one
can then expand it to first order in the eccentricity).

To see that there is no scale-independent term,
notice that such a term would have to result from
integrating an entropy density with units of $1/L$.
Such terms are of the form:
 \begin{equation}
F(s)=f(\frac{dx}{ds},\frac{dy}{ds}) (\frac{d^2x}{ds^2})
+g(\frac{dx}{ds},\frac{dy}{ds}) \frac{d^2y}{ds^2},
\end{equation}
where $s$ is the arc-length.

These terms are total derivatives: let $\alpha$ define
the angle of the tangent vector.  Then $\frac{dx}{ds}=\cos \alpha,
\frac{dy}{ds}=\sin \alpha$. The term equals
$[-f(\cos \alpha, \sin \alpha) \sin \alpha+g(\cos \alpha,\sin \alpha) \cos
\alpha]  \frac{d \alpha}{ds}$.

Now the total entropy, $\int F(s)ds$
can be rewritten as an integral with respect to $\alpha$:
\begin{equation}
\int [-f(\cos \alpha, \sin \alpha) \sin \alpha+g(\cos \alpha,\sin \alpha)
\cos \alpha] d \alpha
\end{equation}
By the symmetry $s \rightarrow -s$, $f$ and $g$ must be even functions of cosine and sine.
Therefore the integral is
equal to zero since the contributions from $\alpha$ and $\alpha+\pi$
cancel one another.

While there are no constant terms for smooth regions in
two dimensions, shapes with corners do have scale-independent
terms that can be attributed to the corners
(this can happen even if the rotational symmetry is not broken,
as shown for the quantum Hall state in Ref. \cite{sierra2010} )

The first anomalous term in two dimensions scales as $\frac{1}{L^2}$;
one example of such a term is
\begin{equation}
\int ds \left(\frac{\partial^2 x}{\partial s^2}\right)^3.
\end{equation}
This expression does not vanish identically,
as can be seen by an example of a region which is nearly circular,
i.e., described by the polar coordinates
\begin{equation}
r=1+\epsilon(\theta),
\label{eq:fingers}
\end{equation}
where $\epsilon\ll 1$. Evaluating the integral to linear order
in $\epsilon$ one obtains
\begin{equation}
\int ds\left(\frac{\partial^2 x}{\partial s^2}\right)^3\approx
-\int ds(\cos^3\theta)(1-6\epsilon(\theta)-3\epsilon'(\theta)\tan\theta)
\end{equation}
which is nonzero for $\epsilon(\theta)\propto \cos 3\theta$.

In higher dimensions, the entropy can have terms
that depend on
$\frac{\partial x^i}{\partial u^\alpha}$ and
higher derivatives. When rotational symmetry is broken,
the Cartesian indices do not have to be contracted but the indices for the
coordinates on the surface still do, because the entropy
has to be independent of the coordinate system. The first allowed
correction to the entropy scales as $\frac{L^{D-1}}{L}$ and a
simple term in the entropy density that leads to such a correction
\begin{equation}
\left(\frac{D^2x}{Du^\alpha Du^\beta}g^{\alpha\beta}\right)\left(\frac{\partial x}{\partial
u^\gamma}\frac{\partial x}{\partial u^\delta}g^{\gamma\delta}\right) \label{eq:minusone}
\end{equation}
The first factor gives the proper scaling, $\frac{1}{L}$.
 The other
factors ensure that the expression is not a total derivative
in dimensions above two. (The simplest term that
scales as $\frac{1}{L}$, $\frac{D^2 x}{Du^\alpha Du^\beta}g^{\alpha\beta}=\nabla^2 x$
is a total derivative.)
The indices are all contracted in pairs so the expression is
coordinate-invariant.

Once a term of order $L^{D-2}$ has appeared,
one would expect that all terms of lower orders appear too,
and that is what one finds.  Since the $L^0$
term is especially important, we checked explicitly
that in even dimensions greater than two, a scale-invariant term $L^0$
is allowed as long as there is no symmetry.  As an example,
take Eq. (\ref{eq:minusone}) and multiply it by
a power of the mean curvature,
$\left(g^{\alpha\beta}\kappa_{\alpha\beta}\right)^{D-2}$. This has the units
$L^{1-D}$, therefore it gives a scale-invariant contribution when integrated (note that since $D$ is even, there
are an even number of powers of $\kappa$ as required). Both Eq. (\ref{eq:minusone})
and this term with the extra factors of $\kappa$
give a nonzero entropy for a generic region.  As an example, consider
the $D$-dimensional surface of revolution obtained by rotating Eq.
(\ref{eq:fingers}) (which can be regarded as a curve in the $x_1,x_2$
plane in $D$-dimensions)
around the $x_1$-axis (explicitly,
$x_1=r(\theta)\cos\theta,\sqrt{(x_2)^2+\dots+(x^{D})^2}=r(\theta)\sin\theta$). Both
integrals
are nonzero for $\epsilon(\theta)\propto\cos 3\theta.$

\subsection{Broken Parity Symmetry} \label{sec:par}
Breaking just parity can
also lead to terms whose exponents deviate from Eqn.\ref{expanL}
even if rotational symmetry still exists, when the dimension is odd.
However, these additional terms all vary as a negative power of $L$. In even
dimensions, parity does not lead to any additional terms.

The extra terms
arise from an additional tensor, $\gamma$, that
is allowed when parity is broken. This
tensor is related to the antisymmetric tensor $\epsilon$:
\begin{equation}
\gamma^{\alpha_1\dots\alpha_{D-1}}=\frac{1}{\sqrt{g}}\epsilon^{\alpha_1\alpha_2\dots\alpha_{D-1}}
\end{equation}
where $\sqrt{g}$ is
the square root of the determinant of the metric.  (This coefficient
is necessary for ensuring that $\gamma$ transforms as a tensor).

\begin{figure}
\includegraphics[width=.45\textwidth]{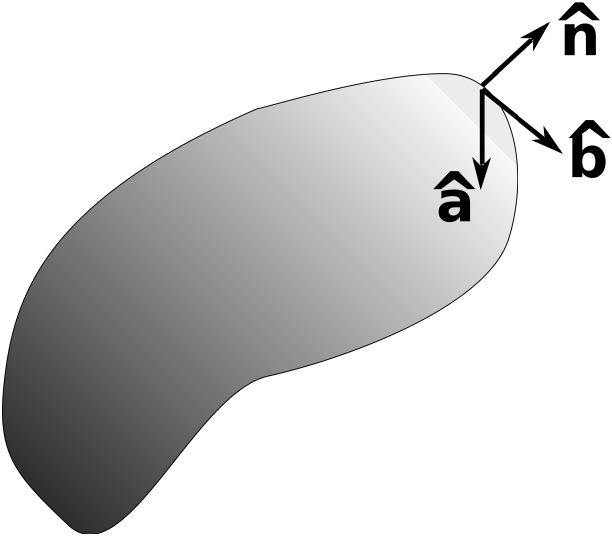}
\caption{\label{fig:swing} Defining the orientation of a hypersurface from
an orientation of space,
illustrated in three dimensions.
A pair of axes on the surface $\hat{a},\hat{b}$ is defined to be
right-handed if the triad $\hat{a},\hat{b},\hat{n}$ is right-handed.
Formally speaking,
$\gamma$ is defined by contracting the D-dimensional epsilon tensor with
the normal $\hat{n}$ and then transforming to curvilinear
coordinates,
$\gamma^{\alpha_1\alpha_2\dots\alpha_{D-1}}=n^{i_d}\epsilon_{i_1i_2\dots
i_d}\frac{\partial x^{i_1}}{\partial u^{\beta_1}}\dots\frac{\partial
x^{i_{D-1}}}{\partial
u^{\beta_{D-1}}}g^{\alpha_1\beta_1}g^{\alpha_2\beta_2}\dots
g^{\alpha_{D-1}\beta_{D-1}}$.}
\end{figure}

To understand these results, first note that the $\gamma$ tensor, like the $\kappa$
tensor, depends on how one chooses the normal to the surface.  The
sign of $\gamma$ depends on how the orientation of the surface is chosen,
and this in turn depends not only on the orientation of space (which is
determined by the parity-violating ground state) but also on
the normal to the surface, see figure \ref{fig:swing}.
Since $\gamma$ is odd under changing the sign of $\hat{n}$, if a factor of
$\gamma$ appears
in the entropy-density, an \emph{odd} number $n_\kappa$ of factors of $\kappa$ must
appear as well
\begin{equation}
n_\kappa\equiv 1\mathrm{(\ mod\ 2)}
\end{equation}

Now the requirement that all the indices of the $\kappa$'s and
its derivatives can be contracted with the upper indices of the $g$'s and the
factor of the $\gamma$ still implies that $n_D$, the number of covariant
derivatives is even if $D$ is odd, because then $\gamma$ has an even
number of upper indices.
So a term that includes a factor of $\gamma$ has units of $L^{-n_\kappa-n_D}$ which
is an \emph{odd}
power of $\frac{1}{L}$.

In an even number of dimensions, the entropy still goes down by two powers
of $L$ at
a time because in this case, there must be an odd number of derivatives as well as
an odd number
of factors of $\kappa$ to respect both the rotational symmetry and the
$Z_2$ symmetry
between the inside and the outside of the region.

The first anomalous term in the entropy is however very small, and scales
as $\frac{1}{L}$
or $\frac{1}{L^3}$ depending on whether the dimension of space is $1$
or $3$ modulo 4 respectively.
All the terms between the area law term, $L^{D-1}$ and the constant term go
in steps of $L^2$. This is essentially because the anomalous terms include
factors of $\gamma$ which has $D-1$ upper indices which
all need to be contracted with something, forcing the term to have at
least $D$ factors
of $\kappa$ or covariant derivatives. (one might think
that $\frac{D}{2}$ factors of $\kappa$  should be enough
since each $\kappa$ has two indices; however, since
$\kappa$ is symmetric and $\gamma$ is antisymmetric, contracting
both indices of $\kappa$ with $\gamma$ gives $0$.)

For illustration, here are some examples of non-zero terms:

\begin{eqnarray}
I_3&=&\int dA
g^{\sigma\tau}\gamma^{\alpha_1\alpha_2}D_{\alpha_1}\kappa_{\alpha_2\sigma}\partial_\tau(\mathrm{tr\ }\kappa)^2 \nonumber \\
I_5&=&\int dA g^{\sigma_1\sigma_2}\gamma^{\alpha_1\alpha_2\alpha_3\alpha_4}
D_{\alpha_1}\kappa_{\alpha_2\sigma_1}D_{\alpha_3}\kappa_{\alpha_4\sigma_2} \nonumber
\mathrm{tr\ }\kappa
\end{eqnarray}
for $3$ and $5$ dimensions, where $\mathrm{tr\ }\kappa$
is the mean curvature times $D-1$, $g^{\alpha\beta}\kappa_{\alpha\beta}$.
 These both scale as $\frac{1}{L^3}$.
Both expressions can be generalized to higher dimensions by introducing
extra factors of $D_{\alpha_{k}}\kappa_{\alpha_{k+1}\sigma_{\frac{k+1}{2}}}$
and adding factors of $g^{\sigma_i\sigma_j}$
to contract all the $\sigma$'s.

\section{Valid Constructions to Extract Topological entropy in D=3} \label{sec:valid3d}

Let us define $\mathfrak{D}[X(A,B,C)]$ where $X$ denotes some property of a manifolds $A,B,C$ to be

\be
\mathfrak{D}[X(A,B,C)] = X_A + X_B + X_C - X_{AB} - X_{BC} - X_{CA} + X_{ABC} \nonumber \label{deltab0}
\ee
For example, in this notation $\gamma_{topo} = \mathfrak{D}[S(A,B,C)$. We note that there is one obvious constraint for the Eqn. \ref{kitpres} to be useful which is that the number
$\mathfrak{D}[b_0(\partial A, \partial B, \partial C)] \neq 0$. This is because $\gamma_{topo} \propto \mathfrak{D}[b_0(\partial A, \partial B, \partial C)]$ where the proportionality constant
is the universal topological constant associated with the phase of matter.

We claim that in $D = 3$, the Eqn. \ref{kitpres} still holds as long as the following condition is satisfied:

\be
\mathfrak{D}[\chi(\partial A, \partial B, \partial C)] = 0 \label{deltab1}
\ee

where $\chi$ denotes the Euler characteristic. This is because this Eqn. \ref{deltab1} guarantees that any dependence on local curvature cancels out on the right hand side of Eqn. \ref{kitpres}.
Since $\chi = 2 - 2 g$, the above Eqn. may be re-expressed as $\mathfrak{D}[g(\partial A, \partial B, \partial C)] = 1$. Just to illustrate this point, consider the case when
$A \cap B \cap C \neq 0$. Generically, three regions in $D = 3$ would intersect along a line. Therefore, there are two possibilities: the region $C$ wraps around the surface defined by
$A \cap B$ or it doesn't. First consider the former possibility. In this case, as one may easily check that
\bea
 g(\partial (AC)) & = & g(\partial A) + g(\partial C) -1 \nonumber \\
 g(\partial (BC)) & = & g(\partial B) + g(\partial C) -1 \nonumber \\
g(\partial (ABC)) & = & g(\partial (AB)) + g(\partial C) -1 \nonumber
\eea

This implies that $\mathfrak{D}[g(\partial A, \partial B, \partial C)] = 1$ and therefore this is a valid construction. An example is provided by the construction in Fig.\ref{fig:3dtee}a. On the other hand, when $C$
does not wrap around $A \cap B$, then

\bea
 g(\partial (AC)) & = & g(\partial A) + g(\partial C) \nonumber \\
 g(\partial (BC)) & = & g(\partial B) + g(\partial C)  \nonumber \\
g(\partial (ABC)) & = & g(\partial (AB)) + g(\partial C)  \nonumber
\eea

and therefore $\mathfrak{D}[g(\partial A, \partial B, \partial C)] = 0$ which implies that this is an invalid construction.

\section{Entanglement Entropy of layered $Z_2$ topological phases} \label{sec:layer}

Here we briefly discuss the layered topologically ordered states mentioned at the beginning of Sec. \ref{sec:decomp}. These phases would lead to a correction $\Delta S_A = -\gamma_{2D} L_z$, for layering
 perpendicular to the $z$ direction where $\gamma_{2D}$ is the topological entanglement entropy associated with the theory living in each layer. For a generic geometry, the total topological entanglement entropy
 $\gamma$ may be written as

\begin{figure}
 \includegraphics[width=250pt, height=160pt]{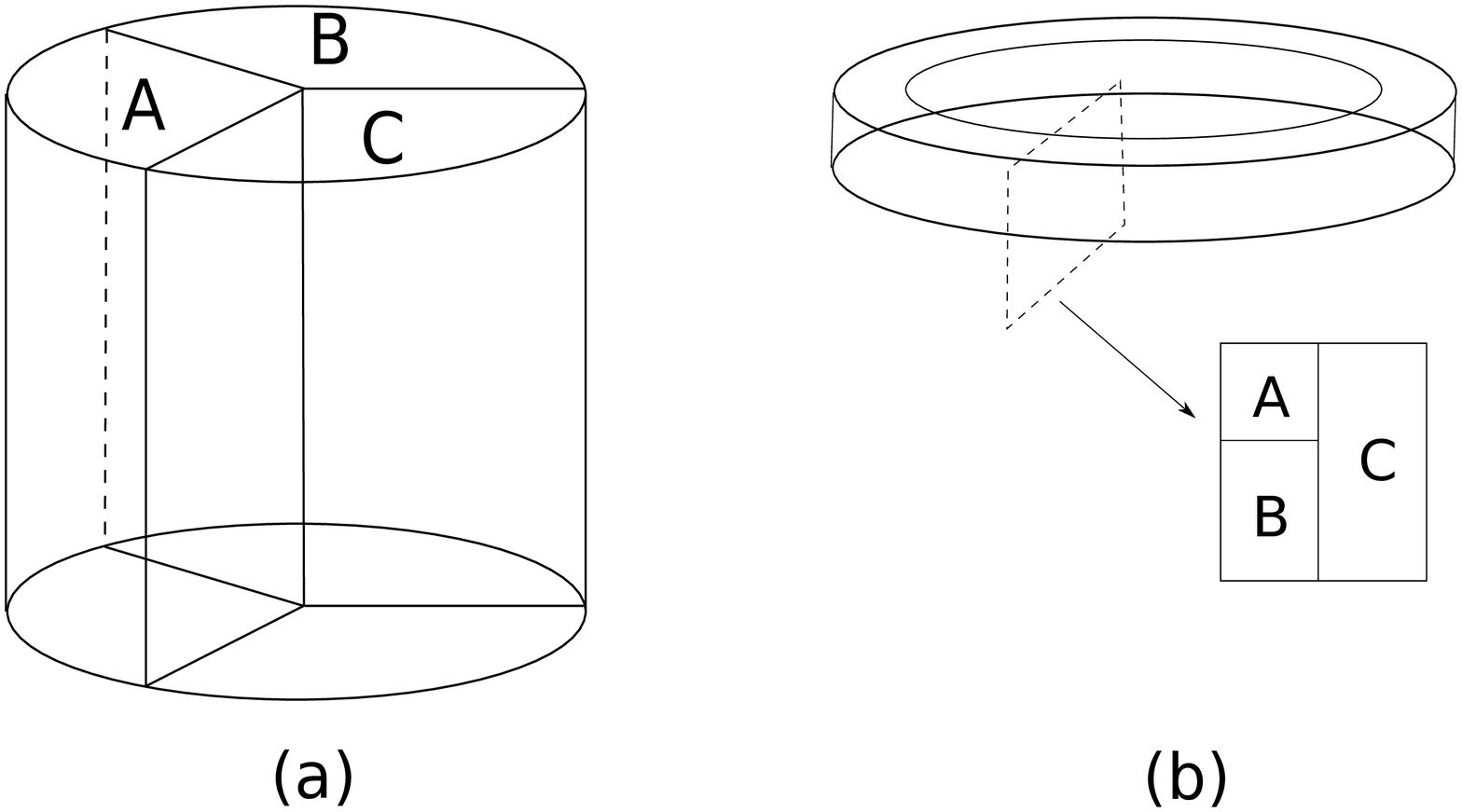}
\caption{Extracting entanglement entropy when a three dimensional topological ordered state coexists with a layered two dimensional topological order. Fig.(a): The $ABC$ construction for this geometry yields
$\gamma = L_z \gamma_{2D}$. Fig.(b): The $ABC$ construction for this geometry yields $\gamma = \gamma_{3d}$.} \label{fig:layer}
\end{figure}

\be
\gamma(L_z) = 2\gamma_{2D}L_z + \gamma_{3D}
\ee

Here $L_z$ is the dimension of regions $A,B,C$ in the $z$ direction. Fig.\ref{fig:layer} shows two different geometries for which the application of $ABC$ formula (Eqn. \ref{kitpres}) yields
$\gamma_{2D}$ and $\gamma_{3D}$ separately.

\section{Linear dependence of $S_{topo}$ on Betti numbers in Three
Dimensions}\label{sec:linear}
Recapitulating the results from the previous section, in  $D = 3$, the topological part of the entanglement entropy is proportional to $b_0$ and is independent of $b_1$ since the genus dependence
can be obtained by patching local Gaussian curvature. Implicit in these statements is an important assumption which is that entanglement entropy depends \textit{linearly} on $b_0, b_1$. This form for the entropy also assumes that the
entropy does not depend on knotting or linking of the toroidal surfaces.
In this section we provide a proof of these two statements. Our only assumption is that the space in which region $A$ is embedded is flat (i.e. has the topology of $\mathbb{R}^3$).

\begin{figure}
\includegraphics[width=.45\textwidth]{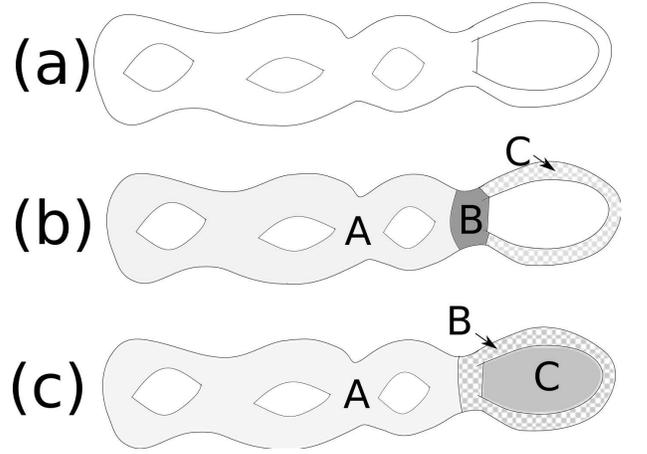}
\caption{\label{fig:tori}
Illustration for proving that the entanglement entropy is linear
in the first Betti number. a) shows a genus $k+1$ torus ($k=3$), and
b) and c) give two ways of dividing it up into $A$, $B$,$C$ regions.
In b) $A$ is a $k$-torus, $B$ is a ball, and $C$ is also
a ball that has been stretched out.
In c) A is a $k$-torus, $B$
a 1-torus, and $C$ a ball filling up the hole of the 1-torus.
Applying strong subadditivity to both configurations gives
a recursion formula for $S(k+1)$ in terms of $S(k)$. Strong
subadditivity gives inequalities, but the left and right
hand sides of the inequality in (b) and (c) are reverses of one another,
leading to an exact relation.}
\end{figure}

\begin{figure}
\includegraphics[width=.45\textwidth]{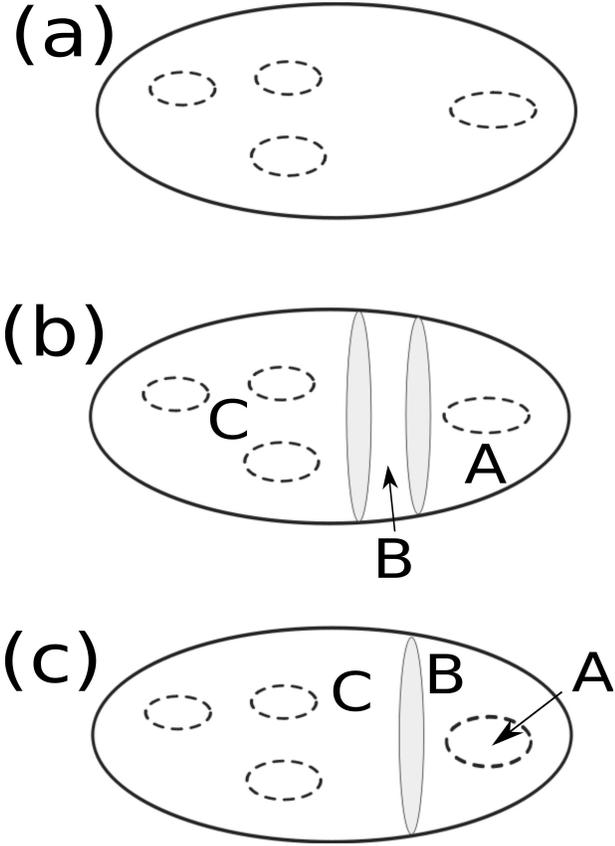}
\caption{\label{fig:spheres} Linear Dependence of the Betti number on
$b_0$.  a) A spherical region with $k+1$ hollow cavities in it ($b_0$
$=k+2$ including the outer surface). b,c)Two sets of regions for applying strong subadditivity
in order to prove that the entropy for $k+1$-holes minus the entropy
for $k$ holes is a constant. b) The region is divided up into $C$,
containing $k-1$ of the cavities, a slab $B$, and a region $A$ containing
the remaining cavity.  c) $C$ again contains $k-1$ of the holes, $B$
is the rest of the region, and $A$ is a solid sphere that fills one of
the cavities.}
\end{figure}

We begin by proving the linearity of the entropy
for the simplest geometries shown in Figs.\ref{fig:tori}a and \ref{fig:spheres}a.
Fig.\ref{fig:tori}a can be used to show that the entropy
is linear in $b_1$.  Consider first the decomposition
shown in Fig.\ref{fig:tori}b
where the boundary of region $A$ is a $k-$torus (i.e. a torus with genus $k$), $B$ and $C$ are three-balls $B^3$ that join together so that the boundary of region $B\cup C$ is
a torus with genus one. We denote by $S_{topo}(k)$ the topological part of the entanglement entropy corresponding to a region whose boundary has genus $k$.
Now apply the strong subadditivity inequality,
$S_{A\cup B} + S_{B\cup C} \geq S_{A\cup B\cup C} + S_{B}$.  The \emph{local} parts
of these entropies satisfy an \emph{equality}, $S_{local}(A\cup B)+S_{local}(B\cup C)=S_{local}(A\cup B\cup C)+S_{local}(B)$ for the configuration
shown, and any configuration where $A$ and $C$
do not meet. The reason is that each patch on the boundaries
of the regions
occurs an equal number of times on both sides of the equation.
Hence the \emph{topological} parts of the entropy also satisfy
\bea
S_{topo}(k) + S_{topo}(1) \geq S_{topo}(k+1) + S_{topo}(0)
\label{eq:upperbhandles}
\eea

Similarly, we construct a different geometry as shown in Fig.\ref{fig:tori}c
where the region $A$ is the same as in Fig.\ref{fig:tori}b, region $B$
has a boundary with genus one, and region $C$ is topologically a
ball $B^3$.
For this topology, adding $C$ to $A\cup B$ \emph{decreases} the number
of
handles from $k+1$ to $k$, so
the strong subaddivity inequality yields the opposite conclusion:

\bea
S_{topo}(k+1) + S_{topo}(0) \geq S_{topo}(k) + S_{topo}(1)
\label{eq:lowerbhandles}
\eea

The above two equations imply that $S_{topo}(k+1) - S_{topo}(k)$ is independent of $k$ and hence $S_{topo}$ is linear in $k$. This
genus dependence can be traded for a non-topological
contribution plus a $b_0$-dependence.

To prove that the topological term is linear in the Betti number $b_0$ ($=$ number of connected components
of the boundary) we repeat the above argument by replacing $k$-tori with a region with $k$
spherical cavities cut in it (see Fig.\ref{fig:spheres}).
This yields
the result that $S_{topo} = \alpha b_0 +  \beta$.
Since $S_{topo}$ should vanish for $b_0 = 0$, this suggests that $S_{topo}$ is strictly proportional to $b_0$.

\begin{figure}
\includegraphics[width=.45\textwidth]{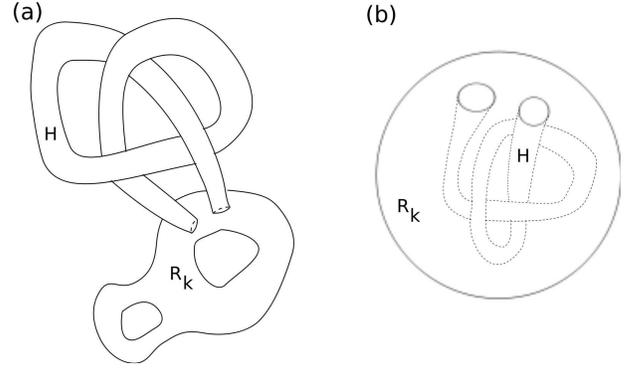}
\caption{\label{fig:handle}Illustration of gluing and drilling.
a) Gluing on a general handle. The torus region is $R_k$, and the shaded region
is $H$, the handle. The handle meets the boundary of region $R$ in a disk, which
is shaded in black. Attaching the handle gives $R_{k+1}=R_k\cup H$.
b)Drilling a hole  The initial region
here, $R_k$, is a solid sphere, and $H$ is a knot that is hollowed
out from the inside of the ball
to obtain $R_{k+1}=R_k-H$. The \emph{boundary} of this region is topologically
a genus one torus, although it is not possible to continuously
deform it into one. Hence drilling \emph{increases}
the genus by one just as adding a handle does.}
\end{figure}

Now we will show that the entropy of \emph{any} three dimensional
region, no matter how knotted, is given by the same formula in terms of
the Betti numbers. First, we assume that $b_0=1$.
The theory of surfaces shows that every connected surface embedded
in three space
is topologically equivalent to one of the $k$-holed tori.
This torus may be built
up by attaching handles repeatedly to increase the genus.
However, there are many ways to attach a handle.
For example, one could add a knotted
handle, like in Fig.\ref{fig:handle}a. The argument in Fig.\ref{fig:tori}c
does not apply to this handle because there is no way to define region $C$--
for the argument above to work it should fill in the hole in the handle so
that the genus decreases by $1$, but this is not possible since the handle
is knotted through itself.

To generalize the argument we therefore construct a proof of the upper bound
that does not require filling the hole in.  We will choose
regions $A$,$B$,$C$ that are all subregions of the initial region
and the handle itself, so that the argument works if there is some linking.

\emph{Ways of Building up a region}
First we will describe all the ways a region whose
boundary is a connected genus $k$ surface can be built up starting from a sphere $R_0$,
so we can be sure that our argument applies to all of them.
Given a region $R_k$ whose boundary is a $k$-torus, one may either glue on a handle $H$
as illustrated in Fig.\ref{fig:handle}a or drill out a hole,
as in Fig.\ref{fig:handle}b. A handle
is a solid cylinder outside the region $R_k$ whose two ends are on the boundary of $R_k$.
Attaching $H$ means defining $R_{k+1}=R_k \cup H$.

At some stages it may be necessary to drill a hole
out of the region instead.
Drilling a hole is an inverted version of the previous process (see Fig.\ref{fig:handle}b):
let $H$ also
have the topology of a cylinder, but $H\subset R_k$, with
its circular faces on the surface of $R_k$.
Then drilling the hole $H$
is passing from $R_k$ to
$R_{k+1}=R_k-H$. For example, applying this process to a ball, as in Fig.\ref{fig:handle}b
gives a ``knot complement," a manifold that is very interesting to knot theorists, since
it encodes the structure of the knot.

We will now prove that $S_{topo}(R_{k+1})-S_{topo}(R_{k})=S_{topo}(torus)-S_{topo}(sphere)$ for
the general case. We
will focus on the case where $R_{k+1}$ is constructed by adding
a handle.  (The other case is similar: if $R_{k+1}$ is obtained by drilling, then the complements, $R_{k+1}^c$
and $R_k^c$, are related by handle-\emph{adding}, so we can apply
our arguments to these complements instead.)

\emph{Change of entropy on adding a handle not affected by links passing through $H$}
We will first show that the entropy changes by $S_{torus}-S_{ball}$
even if the handle $H$ is linked with other portions of $R_k$. In this
argument, we assume that the handle is \emph{not} knotted with
itself; the next argument shows that knotting does not affect the entropy.

\begin{figure}
\includegraphics[width=.45\textwidth]{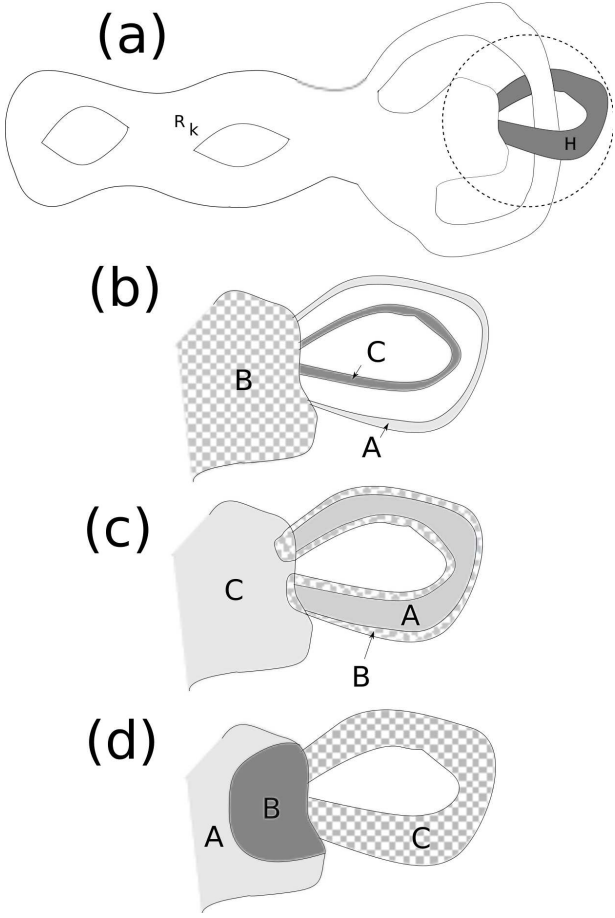}
\caption{\label{fig:slipknot} Figure for the proof that
links passing through the handle do not affect the entropy.
 Part (a) shows the region $R_{k}$ and the handle (in gray) that is added
to it to get $R_{k+1}$.
As in Fig.\ref{fig:tori}, the goal is to relate the entropy of a $k+1$-holed torus to a $k$-holed
one, where $k=3$.
The dotted
circle indicates the portion of the figure that is enlarged
in the later frames of this figure.  The portion of $R_k$ that passes through the hole
is not shown in the subsequent frames for clarity.
Parts (b) and (c) show the two decompositions
that are used
to prove the lower bound on $S_{topo}(R_{k+1})-S_{topo}(R_k)$. In (b), there
are three regions:
Region $B$ (checkered) is $R_k$. The handle is flattened so
that it is ribbon like, and then strip-like regions $A$ and $C$ are
demarcated along its edges.
These together form $A\cup B\cup C:=R_{double}$ which has the topology
of $R_k$ with \emph{two} handles attached.
In (c) we decompose $R_{k+1}$ into $A$,  the center
of the ribbon, $B$ (checkered), the border of the ribbon closed
up with parts of $R_k$ to form a loop, and $C$, the rest of $R_k$.
In this figure $B\cup C=R_{double}$.
Part (D) shows the decomposition used to prove the upper bound, which
corresponds to the division in Fig.\ref{fig:tori}b.}
\end{figure}

Consider the regions in Fig.\ref{fig:handle}a. We will first prove this inequality:
\begin{equation}
S_{topo}(R_{k+1})-S_{topo}(R_k)\geq S_{torus}-S_{ball}
\label{eq:otherhalf}
\end{equation}
where $S_{torus}$ and $S_{ball}$ are the topological
entropies of an unknotted torus and a ball respectively. Begin
by sliding the two ends of the handle (the gray region in Fig.\ref{fig:handle}a)
long the surface of $R_k$
so that they are right next to each other, as shown in the figure.
The proof of this inequality has two steps, using the decompositions illustrated in
Fig.\ref{fig:slipknot}b,c.

First, as in Fig.\ref{fig:slipknot}b, let $B=R_k$, and let $A$ and $C$
be narrow strips
along the left and right side of $H$.
 $A\cup B$ and $C\cup B$ are
two regions that can both be deformed into the same topology ($R_{k+1}$);
altogether $A\cup B\cup C$
forms a region $R_{double}$ with \emph{two} handles attached to $R_k$.
The strong subadditivity implies
\begin{equation}
2S_{topo}(R_{k+1})\geq S_{topo}(R_k)+S_{topo}(R_{double}).
\end{equation}

Next, as in Fig.\ref{fig:slipknot}c, let $A$ be the interior of the strip
$H$. Let $B$ be the border of this strip, marked with a checkered pattern in
the figure. This border is closed  by adding some small
parts of $R_k$ so that it surrounds region $A$ completely. Let $C$ be the rest of $R_k$.
In this decomposition, $C\cup B=R_{double}$, while $A\cup B\cup C=R_{k+1}$,
and $A\cup B$ and $B$ are a ball and a torus respectively.
 Then
\begin{equation}
S_{topo}(ball)+S_{topo}(R_{double}))\geq S_{torus}+S_{topo}(R_k).
\end{equation}
Adding the previous two equations and canceling $R_{double}$
gives Eq. (\ref{eq:otherhalf}).

In this argument it has not been necessary to add on regions \emph{external}
to the handle, so the argument still works if parts of $R_k$ are linked
through it.

Now the reverse inequality
$S_{topo}(R_{k+1})\leq S_{topo}(R_k)+S_{torus}-S_{ball}$
is proved just as in the argument for the simple $k$-tori using
the division shown in Fig.\ref{fig:slipknot}d, which is no different than
the original construction in Fig.\ref{fig:tori}b.
Hence $S_{topo}(R_{k+1})-S_{topo}(R_k)=S_{torus}-S_{ball}$, and
we conclude that $S_{topo}(R_k)=kS_{torus}-(k-1)S_{ball}$;
that is, the entropy is linear in the genus of the surface.

For the case where the surface has more
than one boundary component we just
start with a region whose boundary has many components, which
are all spheres. The entropy of this region is
proportional to $b_0$ by an argument similar to the one illustrated in Fig.\ref{fig:spheres} (this
applies even when the spheres are nested in each other). It is known that \emph{any region may be built
up from such a region by either attaching handles or drilling holes. It is
even possible to choose the handles and holes so that each one
starts and ends on the same component of the boundary
of} $R_k$.  Then the calculation of the entropy as the
handles are attached proceeds just as
above.

\begin{figure}
\includegraphics[width=.45\textwidth]{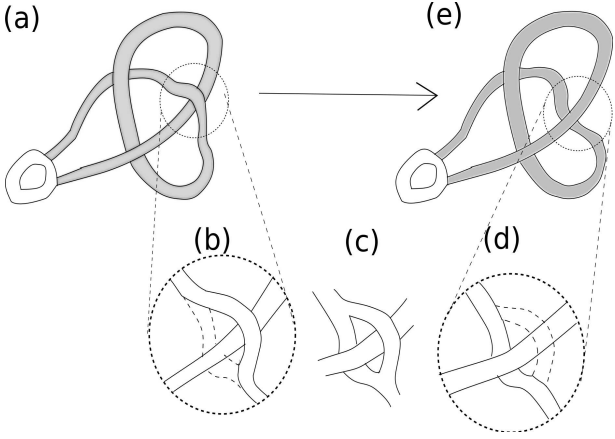}
\caption{\label{fig:ghosts} Showing that the entropy of a knotted handle
is the same as if it were an unknotted.
(a) Shows the knotted handle $H$ (the gray region) that is
attached to the region $R_k$ (the white region). Passing
one strand through the other so the topology changes to figure (e) causes
the handle to become unknotted: the loops in (e) can be
untwisted, giving a simple handle. Any knot can become
unknot-able if the right strands are passed through each other. The
intermediate panels
show that this process  does not change the entropy.
Panel (b) is a blow-up of Fig.(a). In panel (c) the topology is
changed so that one strand passes through an eyelet in the other.  This configuration is obtained
by adding a handle along the dotted lines in
(b), hence the entropy changes by $S_{torus}-S_{sphere}$. (The handle is unknotted)
Panel (d) is obtained by \emph{removing} a handle
from (c)--the entropy therefore decreases back to its original value.
Thus, the change from (a) to (e) does not affect the topological entropy.
}
\end{figure}

\emph{Change of entropy not affected by knotting of the handle}
If the handle is knotted, this argument does not immediately work,
because the entropy change $S_{torus}-S_{ball}$ could be different when
the torus is knotted.
However, a knotted handle may be transformed
to an unknotted one by
repeatedly adding and then removing of \emph{unknotted} handles
as illustrated in Fig.\ref{fig:ghosts}.  By the previous argument,
the entropy returns to its original value after this is done.
Hence the knotted and unknotted handles have the same entropy.

\section{A lightening primer on differential geometry} \label{sec:primer}

We have assumed that the entanglement entropy of a region can be written
as a sum of contributions along the boundary.  These contributions can
depend on the shape of the boundary; for a smooth boundary, the shape is
described, at lowest order, by the curvature tensor.  The purpose of this
section is to describe the curvature tensor and some of the operations
that one can perform on this tensor to get the quantities that the
entanglement entropy can depend upon.

  In two-dimensions, the boundary is a curve, and the curvature of such a
curve at a point $P_0$ is given by
\begin{equation}
-\hat{n}(P_0)\cdot\frac{d^2\mathbf{R}}{ds^2}\label{eq:moonrise}
\end{equation}
where $\hat{n}$ is the normal vector at the point, $\mathbf{R}$
is a vector tracing out the curve and $s$ is the arc-length along
the curve. This formula
can be understood by introducing a new coordinate
system where the $x$-axis is tangent to the surface at $P_0$.  Then the
$y$-value of the curve has a minimum at $x=0$ and the curvature $\kappa$,
intuitively, should describe how rapidly the curve curves away from the
$x$-axis, i.e., $y=H(x)=\frac{1}{2}\kappa x_2+\dots$, which leads
to Eq. \ref{eq:moonrise} because $H(x)$ is given by
$-\hat{n}(P_0)\cdot(\mathbf{R}(x)-\mathbf{R}(x_0))$ where $x$ is used to
parametrize the curve.
(Although the derivative in Eq. (\ref{eq:moonrise}) is taken
with respect to $s$ rather than $x$, the answer is the same because
$x(P)\approx s(P)-s(P_0)$ because the $x$-axis is nearly coincident with
the curve in the vicinity of $P_0$.)

In higher dimensions there are more directions available for
the surface to curve in, so the curvature is given by a tensor.  Again one
begins by rotating the coordinate system of space again so
that it is easy to describe the curvature of the surface and then
rewrites the answer in terms of curvilinear coordinates on
the surface itself. Take the first $D-1$ coordinates
to span a hyperplane tangent to the surface, and let the $D^\mathrm{th}$
coordinate point opposite to the normal vector; the surface
is described by
\begin{equation}
x^D=H(x_1,\dots,x_{D-1})\approx\sum_{i,j=1}^{D-1}
\frac{1}{2}\kappa_{ij}x^ix^j.\label{eq:grid}
\end{equation}
It is convenient not to have to keep redefining the coordinate system for
each point one studies.  One can define $\kappa$ instead
relative to a fixed curvilinear coordinate system on the
surface, described by the coordinates $u^1,u^2,\dots u^{D-1}$.
We obtain
\begin{eqnarray}
\kappa_{\alpha\beta}&=&\sum_{ij}\kappa_{ij}
\frac{\partial x^i}{\partial u^\alpha}\frac{\partial x^j}{\partial
u^\beta}\nonumber\\
&=&\frac{\partial^2 H}{\partial u^\alpha}{\partial u^\beta}\nonumber\\
&=&-\hat{n}\cdot \frac{\partial^2\mathbf{R}}{\partial u^\alpha\partial
u^\beta}.
\end{eqnarray}
The first equation is the transformation rule for covariant
tensors and the second uses Eq. (\ref{eq:grid}) and the chain rule.

In three dimensions, the two ``principle curvatures" $\kappa_1$
and $\kappa_2$
are defined to be the eigenvalues of $\kappa_{ij}$ relative
to the special Cartesian coordinate system.  They are usually
defined more geometrically: let a plane be drawn through the normal and
allow it to sweep through all orientations.  For each orientation, one
obtains an intersection curve and the largest and smallest
curvatures of these curves are the principle curvatures. One
can set up a special coordinate system aligned with the principle
directions since they are orthogonal
to one another.
The
two scalars that are most useful are the mean curvature and
the Gaussian curvature, defined as $\frac{1}{2}(\kappa_1+\kappa_2)$ and
$\kappa_1\kappa_2$.  These expressions can be written for
any Cartesian coordinate system as $\frac{1}{2}\mathrm{tr}\ \kappa$ and
$\mathrm{det}\ \kappa$.  Generalizing these
expressions to curvilinear coordinates cannot be done without
introducing the ``metric."  These expressions do not
apply to the curvilinear coordinates, though, since $\kappa_{curvilinear}=
M\kappa_{Cartesian}M^T$ where $M=\frac{\partial x}{\partial u}$.
Therefore
$\mathrm{tr}\ \kappa_{curvilinear}=\mathrm{tr} M\kappa_{Cartesian} M^T$,
which does not simplify unless $M$ is special.

The technique for making coordinate-invariant expressions is to multiply
tensors with equal numbers of upper and lower indices and to contract them
all. The entropy, for example, cannot depend on the coordinate system
because only the surface itself is important, not how it
is described.  Defining the mean curvature
and Gaussian curvatures for a general coordinate system is a good
illustration of how one makes coordinate-invariant expressions.  For any
$D-1$ dimensional surface there are always two tensors that one can
define.  The curvature is the more complicated of these, and the
other is the \emph{metric} which is the matrix of inner
products of the tangent vectors along the coordinate directions:
\begin{equation}
g_{\alpha\beta}=\frac{\partial\mathbf{R}}{\partial
u^\alpha}\cdot\frac{\partial\mathbf{R}}{\partial u^\beta}
\end{equation}
which describes how stretched or sheared the $u$ coordinates are. If one
does not know the metric $g$, the components of $\kappa_{\alpha\beta}$ do
not mean anything.  They might be very big, but
this could be because the mesh spacing in the $u$-coordinate system is
gigantic.  Instead, one can choose a coordinate system that is
orthonormal according to the metric.  More conveniently, one
usually defines $g^{\alpha\beta}$ to be the inverse of the metric
tensor and then contracts it with $\kappa$ in various ways, to get  scalars
For example, define
\begin{equation}
A=\kappa_{\alpha\beta}g^{\alpha\beta};\ \
B=\kappa_{\alpha\beta}\kappa_{\gamma\delta}g^{\beta\gamma}g^{\alpha\delta}
\end{equation}
These are both independent of the coordinate system, and they
can be calculated by using the orthonormal coordinate system
with the axes along the principle directions.  $A$ and $B$
reduce to $\kappa_1+\kappa_2$ and $\kappa_1^2+\kappa_2^2$.  Hence
$H=\frac{1}{2}A$, $G=\frac{1}{2}(A^2-B)$.
(Rationale for these calculations: The tensor
$C^{\alpha}_\gamma=g^{\alpha\beta}\kappa_{\beta\gamma}$ maps vectors to
other vectors of the same type, $v^\alpha\rightarrow C^{\alpha}_\gamma
v^{\gamma}$ (unlike the original $\kappa$), so its eigenvalues are
well-defined.  One cannot define
the eigenvectors of a matrix that maps one vector space to another,
because there is no way to compare the vectors to see whether
they are proportional to one another.)

The Gauss-Bonnet theorem states,
interestingly, that the
integral of the Gaussian curvature is a topological constant, equal to
$2\pi$ times the
Euler characteristic of the surface; Chern generalized this
result to manifolds of any even dimension:
\begin{equation}
\int_M d^{d-1}A \mathrm{det\ }\kappa=\frac{1}{2}\omega_d \chi(M)
\end{equation}
where $d$ is odd, $\omega_d$ is the area of a sphere in $d$-dimensional
space and $\chi(M)$ is the Euler characteristic of the manifold.
(This result applies to $d-1$ manifolds
that need more than $d$ dimensions to be represented
without self-intersections--it only
has to be rewritten in terms of the
intrinsic curvature $R_{\alpha\beta\gamma\delta}$.)

The last important tool from differential geometry is the covariant
derivative;
it is a generalization of gradients from scalars to vectors.  The most
straightforward way to define a derivative of a vector $\frac{\partial
A^\beta}{\partial u^\alpha}$ does not have any meaning--it measures not
just the variation of the vector, but also
the variation of the coordinate system. For
example, a vector field that does not change along a certain
direction may have changing components because the coordinate axes may bend.
To measure just the variation of the vector, one can first
transform to coordinates that are not rotating--namely,
the Cartesian coordinates of the space, and then differentiate the vector.
After this, one projects
the change in the vector onto the surface, since the component
of the derivative along the normal is determined by the fact
that the surface is curving.  In equations, one defines
\begin{equation}
D_\beta A^\alpha=\sum_{i=1}^{d-1}\frac{\partial u^\alpha}{\partial
x_i}\frac{\partial}{\partial u^\beta} \left( \sum_{\sigma=1}^{d-1}A^\sigma
\frac{\partial x^i}{\partial u^\sigma}\right).
\end{equation}
Although this definition does not make it clear, the covariant derivative
can be calculated without using a coordinate system for the
embedding space; it can be expressed entirely
in terms of the metric $g_{\alpha\beta}$, as is familiar from
general relativity and differential geometry books.


\begin{thebibliography}{x}
\bibitem{nayak2008} C. Nayak, S. Simon, A. Stern, M. Freedman, S. Sarma, Rev. Mod. Phys. 80, 1083 (2008).
\bibitem{kitaev2003} A. Kitaev, Ann. Phys. 303, 2 (2003).
\bibitem{hamma2005} A. Hamma, R. Ionicioiu, and P. Zanardi, Phys. Lett. A 337, 22 (2005); Phys. Rev. A 71, 022315 (2005).
\bibitem{levin2006} M. Levin and X.-G. Wen, Phys. Rev. Lett. 96, 110405 (2006).
\bibitem{kitaev2006} A. Kitaev and J. Preskill, Phys. Rev. Lett. 96, 110404 (2006).
\bibitem{furukawa2007} S. Furukawa and G. Misguich, Phys. Rev. B 75, 214407 (2007).
\bibitem{haque} M. Haque, O. Zozulya and K. Schoutens, Phys. Rev. Lett. 98, 060401 (2007); O. S. Zozulya, M. Haque, K. Schoutens,
and E. H. Rezayi, Phys. Rev. B 76, 125310 (2007); Shiying Dong, Eduardo Fradkin, Robert G. Leigh, Sean Nowling, JHEP 0805:016 (2008); A. M. Lauchli, E. J. Bergholtz and M. Haque, New J. Phys. 12,
075004 (2010).
\bibitem{melko} S. Isakov, M. Hastings, R. Melko, arXiv:1102.1721v1, To appear in Nature Physics (2011)
\bibitem{frank} Y. Zhang, T. Grover, A. Vishwanath, arXiv:1106.0015v3.
\bibitem{chamon2008} C. Castelnovo and C. Chamon,  Phys Rev B 78, 155120 (2008).
\bibitem{footnotebetti} Heuristically, the Betti number $b_k$ measures the dimension of $k$-dimensional closed surfaces within a manifold that
are not boundaries of a $k+1$-dimensional surface. See, e.g., Chapter 3 of Ref. \cite{nakahara} for an introduction.
\bibitem{nakahara} M. Nakahara, \textit{Geometry, Topology, and Physics} (A. Hilger, London, 1990).
\bibitem{bombin2007} H. Bombin  and M. A. Martin-Delgado, Phys. Rev. B 75, 075103 (2007).
\bibitem{dennis2001} Eric Dennis, Alexei Kitaev, Andrew Landahl, John Preskill, J. Math. Phys. 43, 4452-4505 (2002).
\bibitem{hammawen} A. Hamma, P. Zanardi, X.-G. Wen, Phys. Rev. B 72, 035307 (2005).
\bibitem{chaikin} See, e.g., P. M. Chaikin and T. C. Lubensky, Principles of Condensed Matter Physics (Cambridge University Press, Cambridge, 2000).
\bibitem{corner_note} The same symmetry is responsible for relating corner contributions to the entanglement entropy with angles $\theta$ and $2\pi-\theta$\cite{kitaev2006,sierra2010}.
\bibitem{casini2009} H.Casini, M.Huerta, J.Phys.A 42:504007 (2009).
\bibitem{srednicki} M. Srednicki, Phys. Rev. Lett. 71, 666 (1993).
\bibitem{gauss-bonnet} C.B. Allendoerfer, Bull. Am. Math. Soc. 54 , 249 (1948).
\bibitem{sierra2010} Rodriguez and Sierra, J. Stat. Mech. 1012:P12033, 2010.
\bibitem{haah2011} J. Haah, Phys. Rev. A 83, 042330 (2011).
\bibitem{thanklevin} We are grateful to Michael Levin for an illuminating discussion on entanglement entropy of discrete gauge theories.
\bibitem{footnote_degen} Note that in this case there is a ground state degeneracy and the topological entanglement entropy in general will depend on the particular linear combination of the
ground states for which the entanglement is being calculated, see e.g. Ref. \cite{dong}.
\bibitem{dong} Shiying Dong, Eduardo Fradkin, Robert G. Leigh, Sean Nowling, JHEP 0805:016(2008).
\bibitem{lehur2010} H. Francis Song, Stephan Rachel, Karyn Le Hur, Phys. Rev. B 82, 012405 (2010);
H. Francis Song, Christian Flindt, Stephan Rachel, Israel Klich, Karyn Le Hur, Phys. Rev. B 83, 161408(R) (2011).
\bibitem{klich2006} Israel Klich, Gil Refael, Alessandro Silva, Phys. Rev. A 74, 032306 (2006).
\bibitem{levitov2009} Israel Klich, Leonid Levitov, Advances in Theoretical Physics: Landau Memorial Conference, eds. V Lebedev and M V Feigelman; AIP Conference Proceedings, v. 1134, p. 36-45 (2009).
\bibitem{klich2006_2} D. Gioev and I. Klich, Phys. Rev. Lett. 96, 100503 (2006).
\bibitem{helling2011} R. Helling, H. Leschke, and W. Spitzer, Int.Math.Res.Not. 2011:1451-1482, (2011).
\bibitem{swingle2010} B. Swingle, arXiv:1007.4825v1.
\bibitem{calabrese} P. Calabrese, J. Cardy, J. Stat. Mech. 0406:P06002 (2004).

\end{thebibliography}
\end{document}